\documentclass[10pt,aps,prl,twocolumn,nobibnotes,superscriptaddress,nofootinbib,amsmath,amssymb]{revtex4-2}

\usepackage[english]{babel}
\usepackage{CJKutf8}
\usepackage{graphicx}
\usepackage{siunitx}
\usepackage[letterpaper,total={7in,9.5in},top=0.75in,left=0.75in]{geometry}
\usepackage[colorlinks=true,linkcolor=blue,citecolor=blue,urlcolor=blue,filecolor=blue]{hyperref}
\usepackage{tikz}
\usepackage{xcolor}
\usepackage{float}

\definecolor{lime}{HTML}{A6CE39}
\DeclareRobustCommand{\orcidicon}{
	\begin{tikzpicture}
	\draw[lime, fill=lime] (0,0) 
	circle [radius=0.16] 
	node[white] {{\fontfamily{qag}\selectfont \tiny ID}};
	\draw[white, fill=white] (-0.0625,0.095) 
	circle [radius=0.007];
	\end{tikzpicture}
	\hspace{-0.3mm}
}

\foreach \x in {A, ..., Z}{\expandafter\xdef\csname orcid\x\endcsname{\noexpand\href{https://orcid.org/\csname orcidauthor\x\endcsname}
			{\noexpand\orcidicon}}
}

\newcommand{\Sr}[1]{$^{#1}$Sr}
\newcommand{\RedMotTransition}{${^1\mathrm{S}_0} - {^3\mathrm{P}_1}$}
\newcommand{\RedMotLinewidth}{$\SI{7.5}{\kilo\hertz}$}

\newcommand{\BlueMotTransition}{${^1\mathrm{S}_0} - {^1\mathrm{P}_1}$}
\newcommand{\BlueMotLinewidth}{$\SI{30}{\mega\hertz}$}
\newcommand{\TransparencyTransition}{${^3\mathrm{P}_1} - {^3\mathrm{S}_1}$}
\newcommand{\TransparencyLinewidth}{$\SI{3.8}{\mega\hertz}$}
\DeclareSIUnit\gauss{G}

\def\beq{\begin{equation}}
\def\eeq{\end{equation}}
\def\beqa{\begin{eqnarray}}
\def\eeqa{\end{eqnarray}}
\def\bsub{\begin{subequations}}
\def\esub{\end{subequations}}
\def\bal{\begin{align}}
\def\eal{\end{align}}

\begin{document} 

\renewcommand{\figurename}{Fig.}
\renewcommand{\tablename}{Table}

\title{Continuous Bose-Einstein condensation}

\author{Chun-Chia Chen (陳俊嘉)$\orcidA{}$}
\author{Rodrigo Gonz\'{a}lez Escudero$\orcidB{}$}
\affiliation{Van der Waals-Zeeman Institute, Institute of Physics, University of Amsterdam, Science Park 904, 1098XH Amsterdam, The Netherlands}
\author{Ji\v{r}\'{\i} Min\'{a}\v{r}$\orcidC{}$}
\affiliation{Institute for Theoretical Physics, Institute of Physics, University of Amsterdam, Science Park 904, 1098XH Amsterdam, The Netherlands}
\affiliation{QuSoft, Science Park 123, 1098XG Amsterdam, The Netherlands}
\author{Benjamin Pasquiou$\orcidD{}$}
\author{Shayne Bennetts$\orcidE{}$}
\author{Florian Schreck$\orcidF{}$}
\email[]{ContinuousBEC@strontiumBEC.com}
\affiliation{Van der Waals-Zeeman Institute, Institute of Physics, University of Amsterdam, Science Park 904, 1098XH Amsterdam, The Netherlands}
\affiliation{QuSoft, Science Park 123, 1098XG Amsterdam, The Netherlands}

\date{\today}

\begin{abstract}

Bose-Einstein condensates (BECs) are macroscopic coherent matter waves that have revolutionized quantum science and atomic physics. They are essential to quantum simulation~\cite{Bloch2012QSimUltracold} and sensing~\cite{Cronin2009ReviewAtomInterferometry, Bongs2019RevAI}, for example underlying atom interferometers in space~\cite{Lachmann2021AIinSPACE} and ambitious tests of Einstein’s equivalence principle~\cite{Aguilera2014FreeFallUnivAI, Hartwig2015UnivFreeFallRbYb}. The key to dramatically increasing the bandwidth and precision of such matter-wave sensors lies in sustaining a coherent matter wave indefinitely. Here we demonstrate continuous Bose-Einstein condensation by creating a continuous-wave (CW) condensate of strontium atoms that lasts indefinitely. The coherent matter wave is sustained by amplification through Bose-stimulated gain of atoms from a thermal bath. By steadily replenishing this bath while achieving 1000x higher phase-space densities than previous works~\cite{Bennetts2017SSMOTHighPSD, Volchkov2013ContinuousSisyphus}, we maintain the conditions for condensation. This advance overcomes a fundamental limitation of all atomic quantum gas experiments to date: the need to execute several cooling stages time-sequentially. Continuous matter-wave amplification will make possible CW atom lasers, atomic counterparts of CW optical lasers that have become ubiquitous in technology and society. The coherence of such atom lasers will no longer be fundamentally limited by the atom number in a BEC and can ultimately reach the standard quantum limit~\cite{Wiseman1999LasingNoStiEmission, Baker2020HeisenbergAL, Pekker2020LaserLinewidth}. Our development provides a new, hitherto missing piece of atom optics, enabling the construction of continuous coherent matter-wave devices. From infrasound gravitational wave detectors~\cite{Yu2010GravWaveAI, Graham2013GWatomicsensors} to optical clocks~\cite{Chen2005ActiveClockFirstProposal, Meiser2009MilliHzLaser}, the dramatic improvement in coherence, bandwidth and precision now within reach will be decisive in the creation of a new class of quantum sensors.

\end{abstract}

\begin{CJK*}{UTF8}{min}
\maketitle
\end{CJK*}

Interferometers listen to gravitational waves, image the shadow of black holes, and are at the core of countless other high-precision measurements in science and technology. An interferometer achieves its outstanding sensitivity through the interference of waves. For example, the wave-like properties of matter are exploited in the world's most accurate sensors of acceleration and gravity~\cite{Geiger2020ReviewInertialSensor}. Interferometers reach optimal sensitivity and bandwidth by using continuous, coherent, high-brightness sources, which is why many optical interferometers, such as LIGO, are based on CW lasers. However, no continuous matter-wave source with long-range coherence currently exists, which prevents atom interferometers from reaching their full potential. Creating such a CW source of matter will benefit applications ranging from geodesy~\cite{Lisdat2016Geodesy, Grotti2018TransportableClocks}, to tests of Einstein’s equivalence principle~\cite{Aguilera2014FreeFallUnivAI, Hartwig2015UnivFreeFallRbYb}, gravitational-wave detection~\cite{Yu2010GravWaveAI, Graham2013GWatomicsensors, Adamson2018ReportMAGIS, Badurina2020AION, Canuel2018MIGA}, and dark-matter and dark-energy searches~\cite{Safronova2018RevNewPhysicsAtMol, Safronova2019RevVarFundConst}. 

The key to realizing a CW BEC of atoms is to continuously amplify the atomic matter wave while preserving its phase coherence~\cite{Bhongale2000CWLoadingAtLaser}. An amplification process is essential to compensate naturally occurring atom losses, e.g. from molecule formation. It is also needed to replace the atoms that will be coupled out of the BEC for sustaining an atom laser or atom interferometer. Addressing this challenge requires two ingredients: a gain mechanism that amplifies the BEC and a continuous supply of ultracold atoms near quantum degeneracy.

The first steps towards a continuous gain mechanism were taken by~\cite{Chikkatur2002ContinuousBEC}, where merging of independent condensates periodically added atoms to an already existing BEC, but where coherence was not retained across merger events. A Bose-stimulated gain mechanism into a single dominant mode (the BEC) is required to provide gain without sacrificing phase coherence. Such gain mechanisms have been demonstrated using elastic collisions between thermal atoms~\cite{Miesner1998BoseEnhancedBEC, Bhongale2000CWLoadingAtLaser}, stimulated photon emission~\cite{Robins2008PumpedAtomLaser}, four wave mixing~\cite{Deng1999FourWaveMixing, Inouye1999AmplMatterWave} or superradiance~\cite{Schneble2003SuperradiantBEC}. However, in all these demonstrations the gain mechanism could not be sustained indefinitely.

\begin{figure*}[tb]
\includegraphics[width=0.66\textwidth]{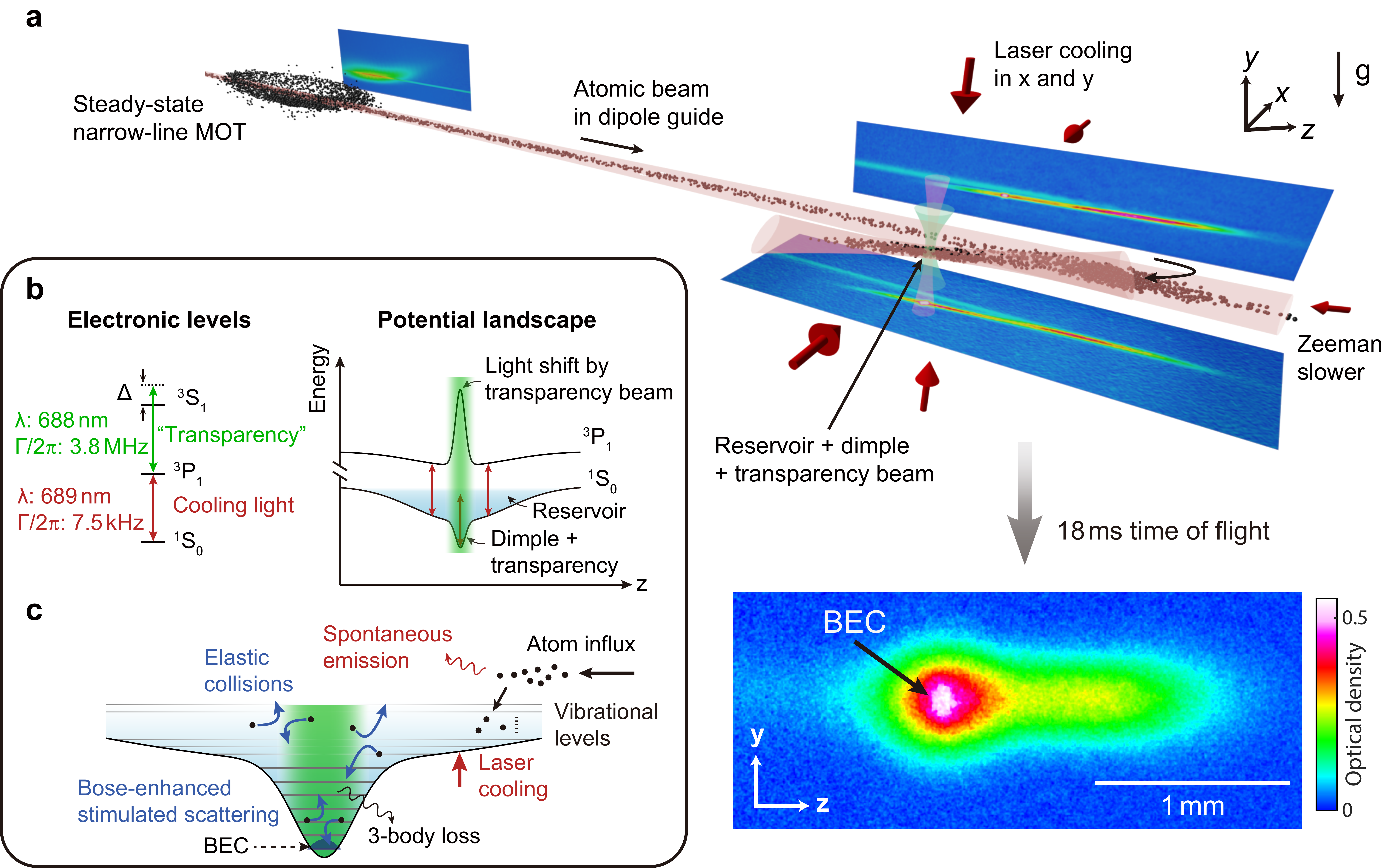}
\caption{\label{fig:Scheme_overview}\textbar \textbf{Experimental setup and scheme.} \textbf{a,} $^{84}$Sr atoms from a steady-state narrow-line magneto-optical trap (MOT) are continuously outcoupled into a guide and loaded into a crossed-beam dipole trap that forms a large reservoir with a small, deep dimple. Atoms accumulate in the laser cooled reservoir and densely populate the dimple, where a Bose-Einstein condensate (BEC) forms in steady state. After time-of-flight expansion, the BEC shows as an elliptical feature in the center of an absorption image. \textbf{b,} By off-resonantly addressing the \TransparencyTransition~transition using a ``transparency" laser beam, we produce a strong spatially varying light shift on the $^3\mathrm{P}_1$ electronic state, rendering atoms locally transparent to laser cooling photons addressing the \RedMotTransition~transition. This enables condensation in the protected dimple region. \textbf{c,} Schematic of the potential landscape from both reservoir and dimple trap, and of the dominant mechanisms leading to BEC atom gain and loss.}
\end{figure*}

To sustain gain the second ingredient is needed: a continuous supply of ultracold, dense gas with a phase-space density --- the occupancy of the lowest motional quantum state --- approaching $\rho = 1$. Great efforts were spent developing continuously cooled beams of atoms~\cite{Lahaye2005EvapBeam, Chen2019Beam} and continuously loaded traps~\cite{Volchkov2013ContinuousSisyphus, Bennetts2017SSMOTHighPSD}, which so far have reached phase-space densities of $\rho = 10^{-3}$. To achieve the required $\si{\micro\kelvin}$ temperatures, these experiments have to use laser cooling, but near-resonant laser cooling light is highly destructive for BECs~\cite{Castin1998LightReabs}. Several experiments have maintained a BEC in the presence of harmful light, either by spatially separating the laser cooling from the quantum gas~\cite{Chikkatur2002ContinuousBEC, Williams1998TheoSSBEC, Lahaye2005EvapBeam, Bennetts2017SSMOTHighPSD, Chen2019Beam} or by reducing the quantum gas' absorbance~\cite{Bhongale2000CWLoadingAtLaser, Cirac1996CollLaserCool, Stellmer2013LaserCoolingToBEC, Urvoy2019DTLaserCoolingBEC}.

Here we demonstrate the creation of a CW BEC that can last indefinitely. Our experiment comprises both ingredients, gain and continuous supply, as illustrated in Fig.~\ref{fig:Scheme_overview}. The centerpiece of the experiment consists of a large ``reservoir" that is continuously loaded with Sr atoms and that contains a small and deep ``dimple" trap in which the BEC is created. The gas in the reservoir is continuously laser-cooled, and exchanges atoms and heat with the dimple gas. A ``transparency" beam renders atoms in the dimple transparent to harmful laser-cooling photons. The dimple increases the gas density while the temperature is kept low by thermal contact with the reservoir. This enhances the phase-space density, leading to the formation of a BEC. Bose-stimulated elastic collisions continuously scatter atoms into the BEC mode, providing the gain necessary to sustain it indefinitely.

\medskip
\noindent
\textbf{Experiment}

\noindent
In order to continuously refill the reservoir, a stream of atoms from an $\SI{850}{\kelvin}$-hot oven flows through a series of spatially-separated laser cooling stages. The initial stages load a steady-state magneto-optical trap (MOT) operated on the \RedMotLinewidth-narrow \RedMotTransition~transition \cite{Bennetts2017SSMOTHighPSD}, shown in Fig.~\ref{fig:Scheme_overview}a. An atomic beam of $\si{\micro \kelvin}$-cold atoms is then outcoupled and guided \cite{Chen2019Beam} $\SI{37}{\milli \meter}$ to the reservoir. This long-distance transfer prevents heating of the atoms in the reservoir by laser cooling light used in earlier cooling stages.

To slow the {$\sim 10$-$\si{\centi \meter \per \second}$} atomic beam and load it into the reservoir while minimizing resonant light, we implement a Zeeman slower on the \RedMotTransition~$|{m_{\mathrm{J}}'=-1}\rangle$ transition. This slower employs a single, counterpropagating laser beam together with the $\SI{0.23}{\gauss \per \centi \meter}$ MOT magnetic field gradient along the guide~\cite{SupplInf}. The $\SI{11.5}{\micro \kelvin}$-deep reservoir is produced by a horizontal {$1070$-$\si{\nano \meter}$} laser beam focused to an elliptical spot with waists $w_y = \SI{14.5}{\micro \meter}$ vertically and $w_x = \SI{110}{\micro \meter}$ horizontally. A $\SI{6}{\degree}$ horizontal angle between the guide and reservoir allows the decelerated atoms to be nudged into the reservoir after reaching the intersection. The atomic beam and the reservoir are radially cooled by two pairs of beams addressing the magnetically insensitive ${^1\mathrm{S}_0} |{m_{\mathrm{J}}=0}\rangle - {^3\mathrm{P}_1} |{m_{\mathrm{J}}'=0}\rangle$ transition.

\begin{figure*}[tb]
\centerline{\includegraphics[width=0.66\textwidth]{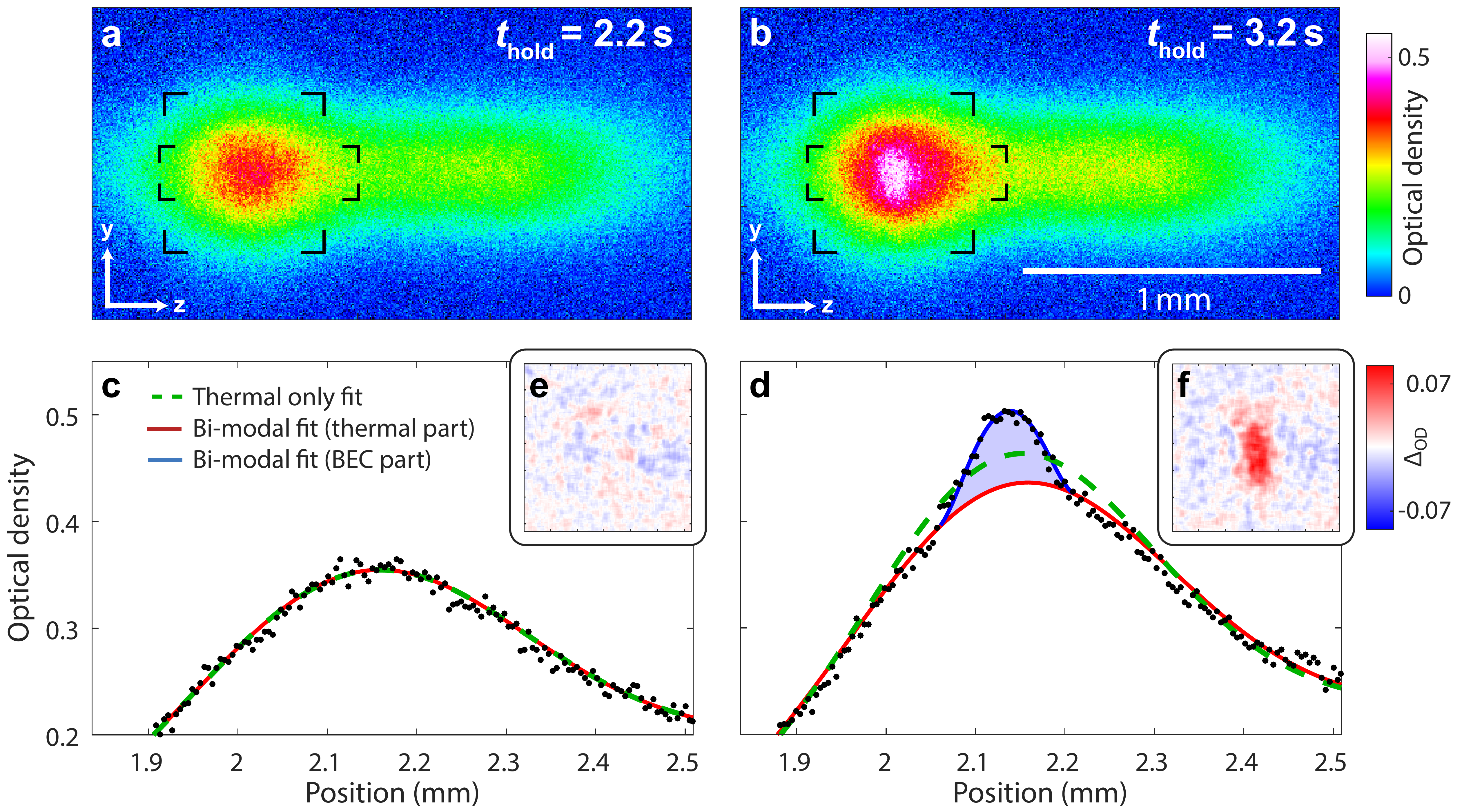}}
\caption{\label{fig:BEC_proof}\textbar \textbf{Detection of the CW BEC.} \textbf{a}, \textbf{b}, Absorption images of the atomic cloud before and after condensation. The atoms are imaged after an $\SI{18}{\milli \second}$ time-of-flight expansion. \textbf{c}, \textbf{d}, Optical density within the rectangles marked by corners in \textbf{a} and \textbf{b}, averaged along $y$. Fitted profiles using a thermal-only distribution (green dashed line) or a bi-modal distribution, consisting of a thermal (red line) and a Thomas-Fermi (blue line) component. The thermal-only fit fails to represent the condensed atoms in \textbf{d} (blue shaded area). \textbf{e}, \textbf{f}, Corner-marked square region of absorption images \textbf{a} and \textbf{b} minus thermal parts of the bi-modal fits, showing the continuous-wave BEC.}
\end{figure*}

This arrangement of traps and cooling beams leads to  the irreversible loading of the reservoir with a flux $\Phi_{\mathrm{R}} = 1.4(2) \times 10^6 \, \mathrm{atoms}\, \si{\per \second}$~\cite{SupplInf}, a radial temperature of $T_{Rr} = \SI{0.85(7)}{\micro \kelvin}$, and an axial temperature of $T_{Rz} = \SI{3.0(5)}{\micro \kelvin}$. The corresponding phase-space flux is $ \kappa = \left( \frac{\partial  \rho_\mathrm{R}}{\partial t} \right)_T = \Phi_\mathrm{R} \left(\frac{\hbar^3 \omega_{Rx} \omega_{Ry} \omega_{Rz}}{k_B^3 T_{Rr}^2 T_{Rz}}\right) = 5.0(2) \times 10^{-2} \, \si{\per \second}$ \cite{Dennis2012TheoCWAtomLaser}, where $\hbar$ is the reduced Planck constant, $k_B$ the Boltzmann constant, and $\omega_{Ri}/2 \pi$ are the reservoir trap frequencies.

To reduce heating and loss we use a ``transparency" laser beam \cite{Stellmer2013LaserCoolingToBEC} that renders atoms in the dimple trap transparent to near-resonant cooling light. This beam is overlapped with the dimple and its frequency is set $\SI{33}{\giga \hertz}$ blue detuned from the \TransparencyTransition~transition, so as to locally apply a differential light shift on the \RedMotTransition~transition, see Fig.~\ref{fig:Scheme_overview}b and \cite{SupplInf}. All transitions to the $^3\mathrm{P}_1$ manifold are thereby shifted by more than $500$ times the \RedMotTransition~linewidth, while atoms in the $^1\mathrm{S}_0$ ground state experience a light shift of only $\SI{20}{\kilo \hertz}$. Without the transparency beam, the lifetime of a pure BEC in the dimple is shorter than $\SI{40}{\milli \second}$ while with the transparency beam it exceeds $\SI{1.5}{\second}$~\cite{SupplInf}.

For a BEC to form in the dimple, the ultracold gas must exceed a critical phase-space density of order one. The dimple is produced by a vertically propagating {$1070$-$\si{\nano \meter}$} beam with $\SI{27}{\micro \meter}$ waist focused at the center of the reservoir. In steady state the $6.9(4) \times 10^5$ atoms in the dimple are maintained at a low temperature ($T_D = \SI{1.08(3)}{\micro \kelvin}$) by thermalization through collisions with the $7.3(1.8) \times 10^5$ laser-cooled atoms in the reservoir \cite{Stellmer2013LaserCoolingToBEC}. The dimple provides a local density boost, thanks to its increased depth ($\SI{7}{\micro \kelvin}$) and small volume compared to the reservoir \cite{Pinkse1997AdiabaticPSD, StamperKurn1998ReversibleBEC, Dutta2015BECgrowthDimple}. This leads to a sufficient phase-space density for condensation. 

In a typical instance of our experiment we suddenly switch all laser beams on and let atoms accumulate in the reservoir and dimple for a time $t_{\rm hold}$. The phase-space density in the dimple increases and eventually a BEC forms. The BEC grows thanks to preferential Bose-stimulated scattering of non-condensed atoms into the macroscopically populated BEC mode. This produces continuous matter-wave amplification, the gain mechanism for the CW BEC \cite{Miesner1998BoseEnhancedBEC}. The BEC grows until losses eventually balance gain and steady state is reached.

\medskip
\noindent
\textbf{Analysis of the CW BEC}

\noindent
We now demonstrate the existence of a BEC and later show that it persists indefinitely. To tackle the first point we analyze atomic cloud density images for $t_{\mathrm{hold}}=\SI{2.2}{\second}$ and $\SI{3.2}{\second}$, immediately before and  after the formation of a BEC, as shown in Fig.~\ref{fig:BEC_proof}a, b. These $x$-integrated absorption images are taken after switching off all laser beams and letting the cloud expand for $\SI{18}{\milli \second}$. Both images show broad distributions of thermal atoms that are horizontally extended, reflecting the spatial distribution of the gas before expansion. Interestingly, the image for the longer $t_{\rm hold}$ shows a small additional elliptical feature at the location of highest optical density, which is consistent with the presence of a BEC. The appearance of a BEC is clearly revealed in Fig.~\ref{fig:BEC_proof}c, d, showing $y$-integrated density distributions. For short $t_{\rm hold}$ only a broad, thermal distribution exists. However for long $t_{\rm hold}$ a bi-modal distribution appears, the hallmark of a BEC.

\begin{figure*}[tb]
\centerline{\includegraphics[width=0.66\textwidth]{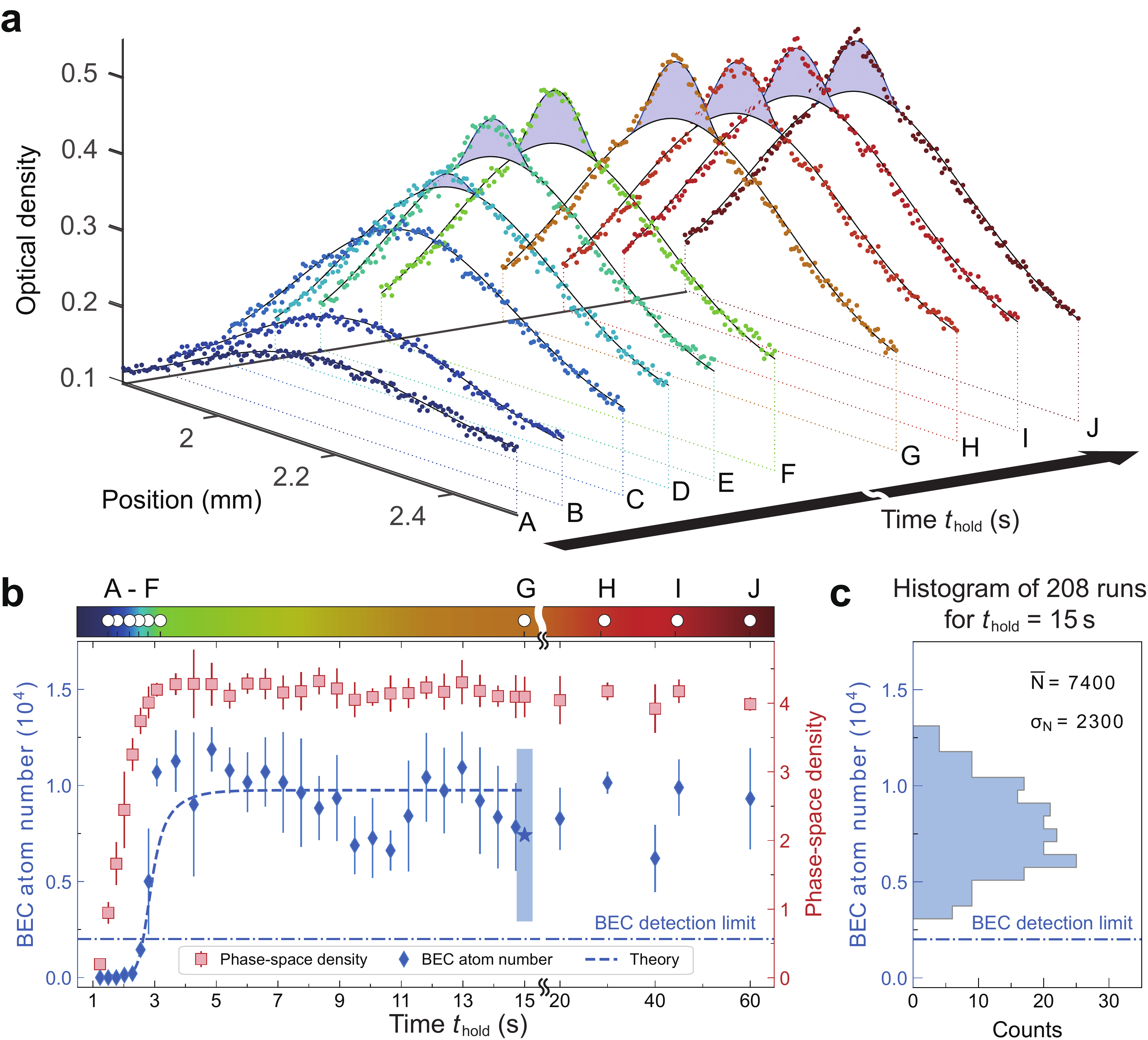}}
\caption{\label{fig:Steady_state_proof}\textbar \textbf{Formation and stability of the CW BEC.} \textbf{a}, Profiles as in Fig.~\ref{fig:BEC_proof}c,d for various hold times (marked in \textbf{b}) first during the formation of the BEC (A: $\SI{1.5}{\second}$, B: $\SI{1.8}{\second}$, C: $\SI{2.2}{\second}$, D: $\SI{2.5}{\second}$, E: $\SI{2.8}{\second}$, F: $\SI{3.2}{\second}$), then during the steady state of the continuous-wave BEC (G: $\SI{15}{\second}$, H: $\SI{30}{\second}$, I: $\SI{45}{\second}$, J: $\SI{60}{\second}$). \textbf{b}, Evolution of the BEC atom number and the dimple atom phase-space density $\rho_D$ in dependence of hold time $t_{\mathrm{hold}}$ after suddenly switching all laser beams on. The dashed blue line shows the result of the BEC evolution fitted to the data before $\SI{15}{\second}$ using the rate-equation model~\cite{SupplInf}. The error bars show the standard deviation from binning $\sim 4$ measurements for each time. \textbf{c}, Histogram of the BEC atom number from $208$ images for $t_{\mathrm{hold}}=\SI{15}{\second}$, long after the establishment of steady state (blue star in \textbf{b}). No points fall below our BEC detection limit of $2000$ atoms. The $\SI{95}{\percent}$ confidence interval ($4 \sigma_{\mathrm N}$) calculated from this data set is given in \textbf{b} at $\SI{15}{\second}$ (blue rectangle).}
\end{figure*}

We further validate the BEC’s existence by fitting theoretical distributions to the absorption images in Fig.~\ref{fig:BEC_proof}a, b. As shown in Fig.~\ref{fig:BEC_proof}c, d, excellent agreement is found by combining a thermal distribution with a Thomas-Fermi (TF) distribution describing the BEC. At short hold times, we find that a thermal fit alone is sufficient to describe the data, whereas at longer times the additional TF component is required, indicating the presence of a BEC. To clearly visualize the BEC, we remove the thermal fit component from the data, see Fig.~\ref{fig:BEC_proof}e, f. The pronounced anisotropic shape of the BEC in Fig.~\ref{fig:BEC_proof}f is consistent with the expansion of a BEC from the anisotropic dimple, whose trap frequency along the $y$ axis is approximately double that along $z$~\cite{SupplInf}.

Once established, the BEC can be maintained in steady state indefinitely with gain balancing losses. As shown in Fig.~\ref{fig:Steady_state_proof}, we study the formation transient and stability of the condensate by recording and analyzing images for different $t_{\mathrm{hold}}$. Fig.~\ref{fig:Steady_state_proof}a shows representative density profiles during the initial $\SI{5}{\second}$ formation transient (A-F) and then in the presence of a stable BEC (G-J). Likewise, Fig.~\ref{fig:Steady_state_proof}b shows the evolution and then stability of the BEC atom number and the peak phase-space density in the dimple $\rho_D = N_D \left(\frac{\hbar^3 \omega_{Dx} \omega_{Dy} \omega_{Dz}}{k_B^3 T_{D}^3}\right)$, where $N_D$ is the thermal atom number in the dimple and $\omega_{Di}/2 \pi$ are the dimple trap frequencies. The steady-state BEC is observed over times much longer than both the lifetime of a pure BEC ($1.5 - \SI{3}{\second}$) and the background-gas limited lifetime ($\SI{7}{\second}$)~\cite{SupplInf}. 

Although we do not continuously monitor the continuous-wave BEC, its atom number fluctuations can be estimated from many independent observations. To study these fluctuations we collected $\sim 200$ measurements for $t_{\mathrm{hold}} = \SI{15}{\second}$, significantly longer than both the lifetimes in the system and the formation transient, see Fig.~\ref{fig:Steady_state_proof}c. A BEC atom number of $\bar{N} = 7400(2300)$ is observed, with none of the points falling below our BEC detection threshold of $2000$ atoms~\cite{SupplInf}.

Modelling the formation, growth and stabilization of the BEC provides valuable insights into this new driven-dissipative system. It also provides the gain and loss from the BEC, which are important for practical applications such as producing a continuous-wave atom laser \cite{Robins2008PumpedAtomLaser} and improving matter-wave coherence. We explain the BEC dynamics by fitting a phenomenological rate-equation model to measured temperature and atom numbers. Our analysis covers the condensate formation and perturbations such as disrupting the reservoir loading~\cite{SupplInf}. From this model we estimate a steady-state gain of $2.4(5) \times 10^5 \, \mathrm{atoms}\, \si{\per \second}$ into the BEC, with representative fits shown in Fig.~\ref{fig:Steady_state_proof}b and Fig.~\ref{fig:SM_Loading_unloading_plus_model}. A substantial fraction of this gain could conceivably be translated into an outcoupled flux forming a CW atom laser. We also find that losses in the BEC at steady state are dominated by three-body recombinations with thermal atoms, due to the gas density exceeding $5 \times 10^{14} \,\mathrm{atoms}\, \si{\per \cubic \centi \meter}$. The presence of high, steady in-flux and loss makes our BEC a driven-dissipative system. We confirm this by showing that it is impossible to model the atoms in the trap as a closed system in thermal equilibrium~\cite{SupplInf}. Open driven-dissipative systems like this one are thought to exhibit rich nonequilibrium many-body physics waiting to be explored, such as purity oscillations \cite{Haine2002StabilityCWAtomLas}, behaviors described by new critical exponents \cite{Sieberer2013DynDrivenDissip}, and unusual quantum phases, especially in lower dimensions \cite{He2017DrivenDiss1D}.

\medskip
\noindent
\textbf{Discussion and conclusion}

\noindent
In summary, we have demonstrated continuous Bose-Einstein condensation. The resulting CW BEC can be sustained indefinitely using constant gain provided by a combination of Bose-stimulated scattering and atom refilling with high phase-space flux. Our work paves the way for continuous matter-wave devices, where matter-wave coherence is no longer limited by the atom number and lifetime of a single condensate \cite{Baker2020HeisenbergAL}. In the near future, BEC purity and matter-wave coherence can be improved by enhancing the phase-space flux loading the dimple. A straight-forward option to achieve this is to render the reservoir laser cooling uniform by using a magic-wavelength reservoir trap. Further options include lowering the reservoir temperature by Raman cooling \cite{Urvoy2019DTLaserCoolingBEC} or by adding a continuously operating evaporation stage \cite{Lahaye2005EvapBeam}. Real-time non-destructive detection and feedback \cite{Szigeti2010FeedbackBECandAL} can be used to stabilize the CW BEC atom number to a stability approaching the shot-noise level \cite{Kristensen2019BECFluc} and even beyond, leading to squeezed states \cite{Johnsson2007SqueezedAL} for measurements beyond the standard quantum limit \cite{Baker2020HeisenbergAL}.

Our CW BEC is the matter-wave equivalent of a CW optical laser with fully reflective cavity mirrors. A tantalizing prospect is to add an output coupler to extract a propagating matter-wave. This could be implemented by coherently transferring atoms to an untrapped state and would bring the long-sought CW atom laser finally within reach \cite{Robins2013RevAtomLaser, Dennis2012TheoCWAtomLaser}. This prospect is especially compelling because our CW BEC is made of strontium, the element used in some of today’s best clocks \cite{Beloy2021ClockNetwork18Acc} and the element of choice for future cutting-edge atom interferometers \cite{Yu2010GravWaveAI, Graham2013GWatomicsensors, Adamson2018ReportMAGIS, Canuel2018MIGA, Badurina2020AION, Tino2019SAGE}. Our work will inspire a new class of such quantum sensors.

\begin{acknowledgments}

\medskip \noindent
\textbf{Acknowledgments}
We thank Francesca Fam\'{a}, Sergey Pyatchenkov, Jens Samland, Sheng Zhou, and the workshop of FNWI, especially Jan Kalwij, Tristan van Klingeren, and Sven Koot, for technical assistance. We thank Vincent Barb\'e, Corentin Coulais, Sebastian Diehl, Klaasjan van Druten, David Gu\'{e}ry-Odelin, Wolf von Klitzing, Ben van Linden van den Heuvell, Robert Spreeuw, Josephine Tan, Premjith Thekkeppat, and Jook Walraven for comments on the manuscript. We thank the NWO for funding through Vici grant No. 680-47-619 and the European Research Council (ERC) for funding under Project No. 615117 QuantStro. This project has received funding from the European Union’s Horizon 2020 research and innovation programme under grant agreement No 820404 (iqClock project). B.P. thanks the NWO for funding through Veni grant No. 680-47-438 and C.-C.C. thanks support from the MOE Technologies Incubation Scholarship from the Taiwan Ministry of Education. 

\noindent
\textbf{Author contribution} C.-C.C. and S.B. built the apparatus. C.-C.C., R.G.E., and S.B. performed the investigation and data collection. C.-C.C., B.P. and S.B. analysed the data. J.M. developed the theoretical model. B.P., S.B., and F.S. supervised the project. F.S. acquired funding. All authors contributed to the manuscript. 

\noindent
\textbf{Correspondence and requests for materials} should be addressed to F.S. Raw data and analysis materials used in this research can be found at: https://doi.org/10.21942/uva.16610143.v1

\end{acknowledgments}


%

\onecolumngrid

\newpage

\twocolumngrid

\setcounter{table}{0}
\setcounter{figure}{0}
\renewcommand{\thetable}{S\arabic{table}}
\renewcommand{\thefigure}{S\arabic{figure}}
\renewcommand{\theHtable}{Supplement.\thetable}
\renewcommand{\theHfigure}{Supplement.\thefigure}

\section*{Materials and methods}

\noindent
\textbf{Creating an ultracold \Sr{84} beam}

\noindent
We use the experimental scheme developed in our previous work \cite{Bennetts2017SSMOTHighPSD, Chen2019Beam} to create an ultracold \Sr{84} beam propagating within a dipole trap guide. The scheme begins with strontium atoms emitted by an $\SI{850}{\kelvin}$-hot oven. They then travel through a succession of laser cooling stages arranged along multiple connected vacuum chambers using first the \BlueMotTransition~and then the~\RedMotTransition~transitions. Using the \BlueMotLinewidth-wide \BlueMotTransition~transition is necessary to efficiently slow and cool the fast atoms from the oven. However, this strong transition can't be used in the last chamber where the BEC is located, due to the likely heating of the BEC from scattered near-resonant photons. Cooling using the narrow \RedMotTransition~transition is however made possible in this last chamber thanks to the addition of a transparency beam (see below).

To form a guided beam, atoms arriving in the final vacuum chamber are first captured and cooled in a narrow-line magneto-optical trap (MOT). They are then outcoupled into a long, horizontal dipole guide with a $\SI{92}{\micro \meter}$ waist. The \Sr{84} atoms propagate along the guide with a velocity $v_{\mathrm{G}} = \SI{8.8(8)}{\centi \meter \per \second}$, a Gaussian velocity spread $\Delta v_{\mathrm{G}} = \SI{5.3(2)}{\centi \meter \per \second}$, and a flux $\Phi_{\mathrm{G}} = 8.6(1.0) \times 10^6 \, \mathrm{atoms}\, \si{\per \second}$.

\medskip
\noindent
\textbf{Making the reservoir and dimple traps}

\noindent
The $\SI{11.5}{\micro \kelvin}$-deep reservoir is produced by a right circularly polarized $\SI{1070}{\nano\meter}$ laser beam propagating in the $z$ direction. It uses $\SI{540}{\milli \watt}$ of power focused to an elliptical spot with waists of $w_y = \SI{14.5}{\micro \meter}$ vertically and $w_x = \SI{110}{\micro \meter}$ horizontally. The guided atomic beam and the reservoir intersect with a horizontal angle of $\SI{6}{\degree}$, about $\SI{1}{\milli \meter}$ from the reservoir center and $\SI{37}{\milli \meter}$ from the MOT quadrupole center. The reservoir beam crosses approximately $\SI{45(10)}{\micro\meter}$ below the guide beam and descends with a vertical tilt of around $\SI{1.2(1)}{\degree}$ as it separates from the guide beam. A secondary {250-$\si{\milli \watt}$} beam of $\SI{175(25)}{\micro \meter}$ waist runs parallel to the guide and points at the reservoir region. The fine adjustment of these beams is used to optimize the flow of atoms from the guide to the reservoir.

The dimple region has a $\SI{7}{\micro \kelvin}$ deeper potential located at the center of the reservoir. This is mainly produced by a vertically propagating 1070-nm ``dimple beam", although $\SI{1}{\micro \kelvin}$ is due to the vertically propagating transparency beam. 
The dimple beam uses $\SI{130}{\milli \watt}$ of power linearly polarized along the $z$ axis with a $\SI{27}{\micro \meter}$ waist in the plane of the reservoir. The dimple trap frequencies are $\left( \omega_{Dx}, \, \omega_{Dy}, \, \omega_{Dz} \right)  = 2 \pi \times \left( 330, \, 740, \, 315 \right) \, \si{\hertz}$, whereas the reservoir beam alone produces a trap with frequencies $\left( \omega_{Rx}, \, \omega_{Ry}, \, \omega_{Rz} \right) = 2 \pi \times \left( 95, \, 740, \, 15 \right) \, \si{\hertz}$.

\medskip
\noindent
\textbf{Zeeman slower on the \RedMotTransition~transition}

\noindent
To load the guided atomic beam into the reservoir it must first be slowed and pushed into the reservoir. To perform this task, we implement a Zeeman slower using the \RedMotTransition~transition starting $\sim \SI{3}{\milli \meter}$ before the guide-reservoir intersection. The slower makes use of the quadrupole magnetic field of the narrow-line MOT to provide a magnetic gradient along the guide's axis. The MOT's quadrupole field has gradients of $-0.55$, $0.32$ and $\SI{0.23}{\gauss \per \centi \meter}$ in the $x$, $y$, and $z$ directions respectively. The slower is displaced by $\SI{37}{\milli \meter}$ along the $z$ axis with respect to the quadrupole center, resulting in a magnetic field offset of $\SI{0.85}{\gauss}$. The slower uses a counter-propagating $\SI{200}{\micro \meter}$-waist laser beam that crosses the guide at a shallow horizontal angle of $\SI{4}{\degree}$. We modulate the laser frequency to broaden its effective linewidth to $\SI{50}{\kHz}$. This makes the slowing robust to potential fluctuations in the effective detuning (see Table~\ref{tab:SM_NarrowLinewidthLaserParameters}). The light intensity corresponds to $2.2 \, I_{\mathrm{sat}}$ when not frequency-broadened, where $I_{\mathrm{sat}} \approx \SI{3}{\micro \watt \per \square \centi \meter}$ is the transition's saturation intensity. We choose the laser detuning to match the Zeeman shift of the $^3\mathrm{P}_1$ $| \mathrm{J'} = 1, {m_{\mathrm{J}}'=-1} \rangle$ state at the intersection between the guide and reservoir. This way atoms reach zero axial velocity at the intersection before being pushed back and into the reservoir.

\begin{table*}[tb]
\caption{\textbar Properties of laser beams addressing the narrow-linewidth \RedMotTransition~transition. Under ``Detuning" $\Delta_1 : \delta : \Delta_2$ refers to a comb of lines from detuning $\Delta_1$ to $\Delta_2$ with a spacing of $\delta$, obtained by triangular frequency modulation.}
\label{tab:SM_NarrowLinewidthLaserParameters}
\resizebox{\textwidth}{!}{
\begin{tabular}{ccccc}
\textbf{Beam name} & \textbf{Detuning} & \textbf{Total power} & \textbf{$1/e^2$ radius} & \textbf{Comments}\\
 & \textbf{(MHz)} & \textbf{($\si{\micro \watt}$)} & \textbf{(mm)} & \\
\noalign{\smallskip}\hline\hline\noalign{\smallskip}
MOT X & $-0.66 : 0.015: -2.2$ & $1.2 \times 10^{3}$ & $23.5$ & two counter-propagating beams \\
\hline \noalign{\smallskip}
MOT Y & $-0.96: 0.02: -3.6$ & $11.3 \times 10^{3}$ & $34$ & single beam, upward propagating \\
\hline \noalign{\smallskip}
MOT Z & $-0.825 : 0.017 : -1.25$ & $7 $ & $4$ & two counter-propagating beams \\
\hline \noalign{\smallskip}
Launch & $+0.9: 0.017: -0.25$ & $20 \times 10^{-3}$ & $0.25$ & single beam \\
\hline \noalign{\smallskip}
Zeeman slower & $-1.74 : 0.017 : -1.79$ & $4.5 \times 10^{-3}$ & $0.2$ & single beam \\
\hline \noalign{\smallskip}
Counter Zeeman slower & $-1.77 : 0.017 : -1.79$ & $10.5 \times 10^{-3}$ & $0.15$ & single beam \\
\hline \noalign{\smallskip}
Molasses X & $+0.042$  & $1.5$ & $14.4$ & two counter-propagating beams \\
\hline \noalign{\smallskip}
Molasses Y(up) & $+0.042$  & $3.5$ & $18$ & single beam, upward propagating \\
\hline \noalign{\smallskip}
Molasses Y(down) & $+0.042$ & $160 \times 10^{-3}$ & $19$ &  single beam, downward propagating \\
\end{tabular}
}
\end{table*}

\medskip
\noindent
\textbf{Loading the reservoir}

\noindent
Since the reservoir is a conservative trap, efficiently loading atoms from the guide requires a dissipative mechanism. This is provided in two ways by laser cooling on the \RedMotTransition~transition. The first is a ``counter Zeeman slower" beam propagating approximately along the $z$ axis opposing the Zeeman slower beam. This beam addresses the $^3\mathrm{P}_1$ $| \mathrm{J'} = 1, {m_{\mathrm{J}}'=-1}\rangle$ state with a peak intensity of $\sim 8 \, I_{\mathrm{sat}}$ and has a waist of $\SI{150}{\micro \meter}$. Making use of this magnetic transition, we choose the light detuning such as to address the atoms near the guide-reservoir intersection and thus compensate the backward acceleration of the Zeeman slower beam. This allows atoms to gradually diffuse toward the reservoir center, where collisions and the second laser cooling mechanism will further lower their temperature. 

The second cooling mechanism consists of a molasses on the radial axes $\left(x,y\right)$ addressing the magnetically insensitive $\pi$ transition. Using a magnetically insensitive transition avoids impacting cooling by the spatial inhomogeneities in the effective detuning due to magnetic field variation across the extent of the laser-cooled cloud. Another cause of spatial inhomogeneities, that does affect the molasses cooling efficiency, is the differential light shift induced by the reservoir trap. This shift is around $+\SI{55}{\kilo\hertz}$, many times larger than the transition's linewidth. The optimal molasses cooling frequency is found to be $\SI{42}{\kilo \hertz}$ higher than the unperturbed transition. This partially accommodates for the differential light shifts and preferentially cools atoms located near the bottom of the reservoir. To reach the lowest temperature and enable condensation in the dimple, we also apply a very low total light intensity of $0.4 \, I_{\mathrm{sat}}$. With this choice of detuning and intensity, some of the incoming atoms reach the reservoir center where they are radially cooled to $T_{Rr} = \SI{0.85(7)}{\micro \kelvin}$. Other atoms might be heated out of the $\SI{9}{\micro \kelvin}$-evaporation-threshold trap by the blue detuned light in the outer trap region.

\medskip
\noindent
\textbf{Minimizing heating and loss in the reservoir}

\noindent
The atoms in the reservoir have a lifetime of $\SI{7}{\second}$ limited by collisions with the background gas of the vacuum chamber. However, these losses can be overwhelmed by optical effects such as photo-association or heating by photon scattering. It is therefore critical to minimize the exposure of the reservoir to unnecessary light, and we address this point by implementing four techniques.

\begin{figure}[tb]
\centerline{\includegraphics[width=0.98\columnwidth]{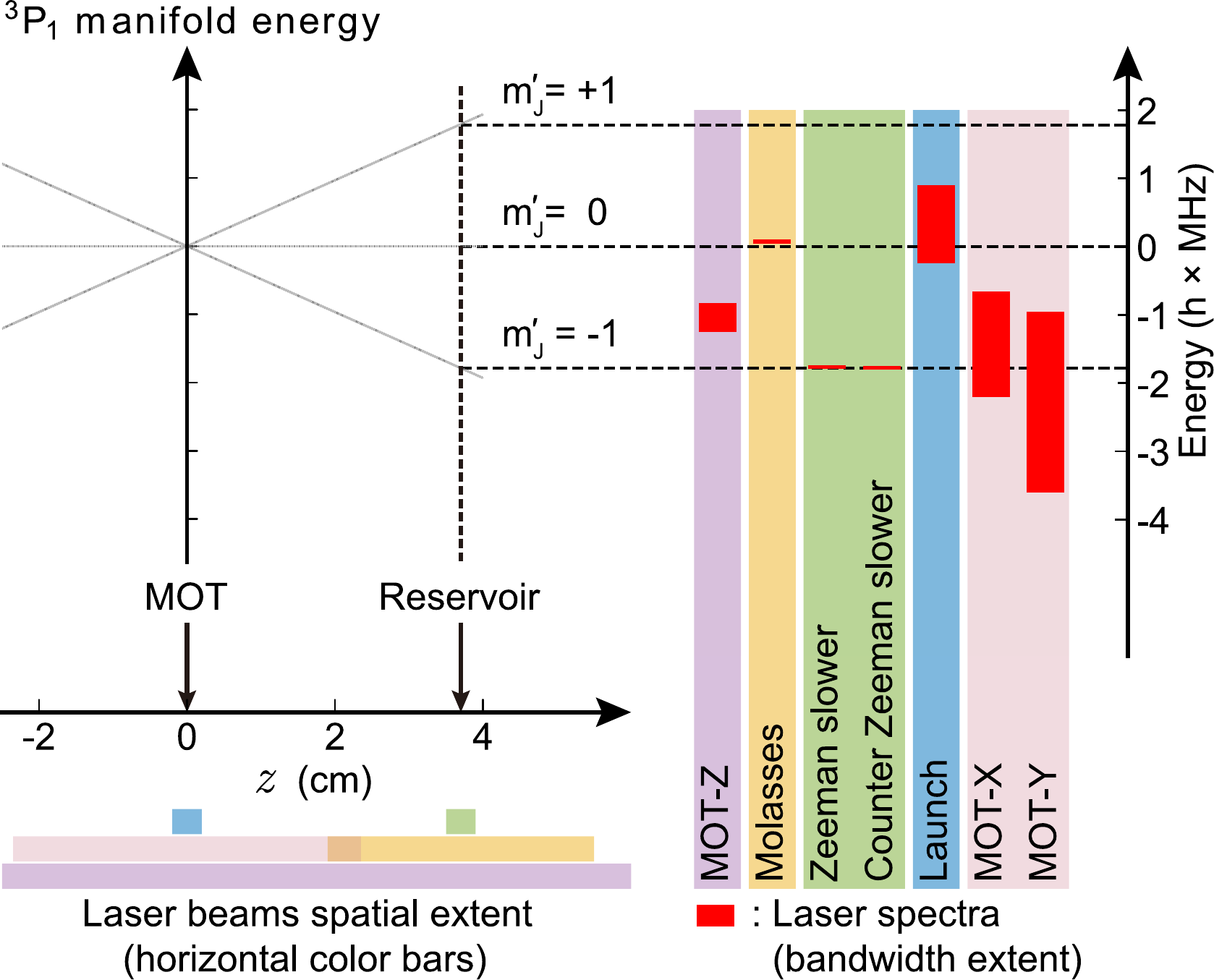}}
\caption{\label{fig:SM_Beams_spectral_spatial}\textbar \textbf{Spectra of narrow-linewidth cooling lasers, and their spatial extent.} The right side represents the spectra of cooling lasers addressing the \RedMotTransition~transition (vertical red bars) with respect to the (relative) energy of the states in the $^3\mathrm{P}_1$ manifold, shown on the left side. The energies of these $m_{\mathrm{J}}'$ states are given depending on the location along the $z$ axis, and the horizontal black dashed lines represent their respective Zeeman shifts when atoms are located inside the reservoir. The horizontal color bars at the bottom left show the location and spatial extent of each laser beam, see also Table~\ref{tab:SM_NarrowLinewidthLaserParameters} for detailed beam parameters.}
\end{figure}

Firstly, the $\SI{37}{\milli \meter}$ offset between the MOT and reservoir centers allows us to avoid any direct illumination from the $x,y$ MOT beams on the reservoir, see Fig.~\ref{fig:SM_Beams_spectral_spatial}. On the $z$ axis, the influence of the MOT beams is greatly reduced by using a ``dark cylinder'', as described in \cite{Chen2019Beam}.

Secondly, we optimize the cooling spectrum and intensity of each laser cooling beam entering the last vacuum chamber. By separately measuring their influence on the reservoir atom number, we optimize on a compromise between the lifetime of atoms and the loading flux. The results are illustrated in Fig.~\ref{fig:SM_Beams_spectral_spatial} and Table~\ref{tab:SM_NarrowLinewidthLaserParameters}. 

Thirdly, we maximize the $\pi$ polarization component of the molasses beams that illuminate both the guided beam and the reservoir, thus minimizing the effects of unwanted transitions. Unavoidably, beams along the $y$ axis possess admixtures of $\sigma^-$ and $\sigma^+$ due to the orientation of the local magnetic field. 

Finally, we purify the spectrum of the light used to address the \RedMotTransition~cooling transition. Our cooling light is produced by multiple injection-locked diode lasers beginning from a single external-cavity diode laser (ECDL). We reduce the linewidth of this ECDL to $\SI{2}{\kilo \hertz}$ by locking it onto a cavity with a finesse of $\sim 15000$, whose spectrum has a full-width half maximum of $\sim \SI{100}{\kilo \hertz}$. By using the light transmitted through this cavity to injection lock a second diode laser, we can filter out the ECDL's amplified spontaneous emission and servo bumps. This filtering is critical to increase the atoms' lifetime inside the dimple by reducing resonant-photon scattering. 

\begin{figure}[tb]
\centerline{\includegraphics[width=0.98\columnwidth]{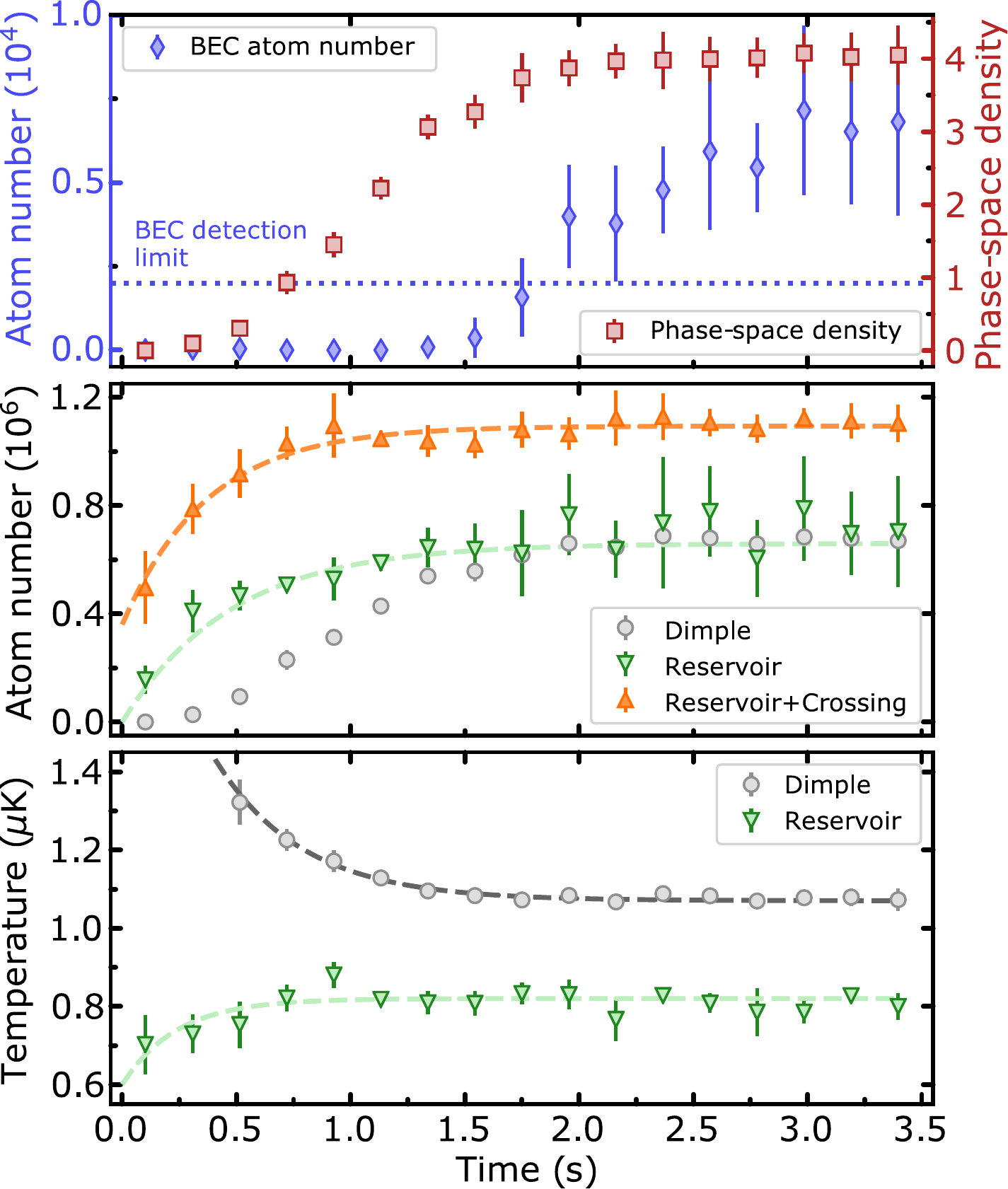}}
\caption{\label{fig:SM_Loading_at_constant_flux}\textbar \textbf{Loading of the reservoir and dimple at constant flux.} We achieve a constant flux $\Phi_{\mathrm{G}}$ in the guide by switching the experiment on for $\SI{10}{\second}$ without the Zeeman slower beam, until reaching a steady flow. We then switch this beam on at time $t=0$. We show (top) the BEC atom number and the phase-space density $\rho_D$ in the dimple. The blue dotted line indicates our BEC detection limit in terms of condensed atom number. We show (middle) the dimple, reservoir, and ``reservoir+crossing'' atom number, and we show (bottom) the temperature $T_D$ in the dimple and the temperature $T_{Ry}$ in the reservoir along the vertical axis. The dashed lines are the results from fits with exponential growth or decay, giving access to the (constant) fluxes, one-body loss rate parameters, and thermalization times (see text). Error bars represent one standard deviation $\sigma$ from binning on average $6$ data points.}
\end{figure}

Without the dimple and transparency beams, individual laser cooling beams reduce the lifetime of atoms in the reservoir to no shorter than $\sim \SI{1.5}{\second}$. With the dimple, transparency and all laser cooling beams on, atoms in the reservoir have a $1/e$ lifetime of $\SI{420(100)}{\milli \second}$, as determined from the fits shown in Fig.~\ref{fig:SM_Loading_at_constant_flux}.

\medskip
\noindent
\textbf{Transparency beam}

\begin{figure}[tb]
\centerline{\includegraphics[width=0.98\columnwidth]{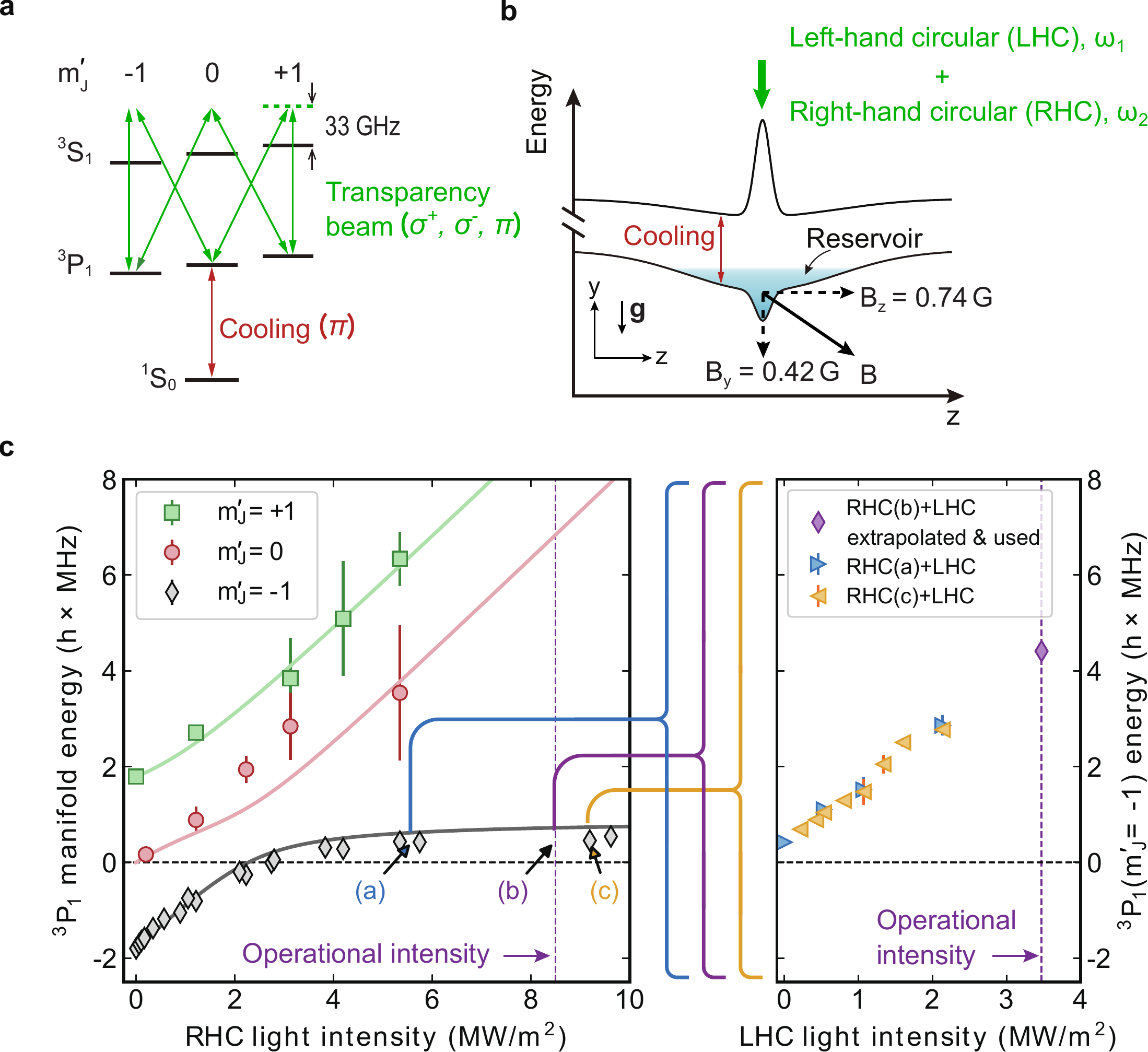}}
\caption{\label{fig:SM_Transparency_effect}\textbar \textbf{Light shift from the transparency beam.} \textbf{a}, Level scheme showing laser cooling and transparency transitions. \textbf{b}, Schematic of the potential energy landscape of reservoir and dimple for the $^1\mathrm{S}_0$ and $^3\mathrm{P}_1$ states. Atoms are rendered insensitive to the laser cooling light by a single vertical “transparency” laser beam (green arrow), containing two frequency components, one for each circular polarization. \textbf{c}, Transition energies to the three $m_{\mathrm{J}}' = 0, \pm 1$ Zeeman sub-levels of the $^3\mathrm{P}_1$ manifold, referenced to the transition at zero electric and magnetic field (black dashed line). The energy shifts are shown for a single right-hand circular (RHC) polarization (left) and with the addition of the left-hand circular (LHC) component (right). We show the solutions (solid lines) of the Schr\"{o}dinger equation for the  ${^3\mathrm{P}_1}$ manifold coupled by a light field with single frequency component and RHC polarization. In this case, at high laser intensities, the energy of the state originating from $m_{\mathrm{J}}' = -1$ saturates, corresponding to the presence of a dark state. The purple vertical dashed lines show the operational intensities of the LHC and RHC light fields used in the CW BEC experiment, and the purple diamond is extrapolated from the data.}
\end{figure}

\noindent
To minimize the destructive effects of resonant light on the BEC and atoms within the dimple, we render this region locally transparent to light on the \RedMotTransition~cooling transition. By coupling light to the \TransparencyTransition~transition, we induce a light shift on the $^3\mathrm{P}_1$ state as illustrated in Fig.~\ref{fig:SM_Transparency_effect}a,b. Due to the extreme sensitivity of the BEC to photon scattering, all sub-levels of the $^3\mathrm{P}_1$ state must be shifted significantly. This requires using at least two of the three transition types $(\sigma^{\pm}, \pi)$ in this $\mathrm{J} = 1 - \mathrm{J'} = 1$ structure. However, when polarizations at the same frequency are combined, quantum interference between sub-levels always produces a dark state in the dressed $^3\mathrm{P}_1$ manifold. In this case the energy of this dark state can only be shifted between $\pm \Delta_{\mathrm{Zeeman}}$, where $\Delta_{\mathrm{Zeeman}}$ is the Zeeman shift of the $^3\mathrm{P}_1$ $m_{\mathrm{J}}'=1$ state. This corresponds to $\Delta_{\mathrm{Zeeman}} = \SI{1.78}{\MHz}$ at the dimple location, giving a light shift that is insufficient to protect the BEC. Thus it is necessary to use different frequencies for the different polarization components of the transparency beam, as illustrated in Fig.~\ref{fig:SM_Transparency_effect}c.

The transparency beam is implemented by a single beam propagating vertically and focused on the dimple location with a $\SI{23}{\micro \meter}$ waist. This geometry aims to minimize the overlap of the transparency beam with the reservoir volume. In this way, we protect atoms at the dimple location without affecting the laser cooling taking place in the surrounding reservoir. This is necessary to maintain the reservoir's high phase-space flux. The transparency laser light is blue detuned by $\SI{33}{\giga \hertz}$ from the \TransparencyLinewidth-wide \TransparencyTransition~transition at $\SI{688}{\nano \meter}$. It contains two frequency components, $\SI{7}{\milli \watt}$ of right-hand circularly polarized light and $\SI{3}{\milli \watt}$ of left-hand circularly polarized light separated by $\SI{1.4}{\giga \hertz}$. The magnetic field at the dimple location lays in the $(y,z)$ plane and has an angle of $\SI{60}{\degree}$ with respect to the vertical $y$ axis along which the transparency beam propagates. This leads to a distribution of the light intensity onto the transitions $\{\sigma^+, \sigma^-, \pi \}$ of $\{1,9,6\}$ for the left-hand and $\{9,1,6\}$ for the right-hand circular polarization.

The light is produced from a single external-cavity diode laser, frequency shifted by acousto-optic modulators and amplified by several injection-locked laser diodes and a tapered amplifier. Since the \RedMotTransition~and \TransparencyTransition~lines are less than $\SI{1.5}{\nano \meter}$ apart, it is crucial to filter the light to prevent amplified spontaneous emission from introducing resonant scattering on the \RedMotTransition~transition. This filtering is performed by a succession of three dispersive prisms (Thorlabs PS853 N-SF11 equilateral prisms), followed by a $\SI{2.5}{\meter}$ (right-hand circular) or $\SI{3.9}{\meter}$ (left-hand circular) propagation distance before aperturing and injection into the final optical fiber. 

\medskip
\noindent
\textbf{Characterizing the transparency beam protection}

\noindent
The transparency beam induced light shifts on the \RedMotTransition~transition were measured spectroscopically by probing the absorption of \Sr{88} samples loaded into the dimple. \Sr{88} is used instead of \Sr{84} since the higher natural abundance improves signal without affecting the induced light shifts. Spectra are recorded for various transparency beam laser intensities at the magnetic field used for the CW BEC experiments. The results are shown in Fig.~\ref{fig:SM_Transparency_effect}c for one then two polarization components.

The observed light shifts are consistent with the calculated dressed states for the six coupled sub-levels of the $^3\mathrm{P}_1$ and $^3\mathrm{S}_1$ states. This is evaluated by solving the Schrödinger equation in the rotating frame of the light field for a transparency beam consisting of a single frequency, right-hand circular laser beam in the presence of the measured external magnetic field. The theoretical results are given in Fig.~\ref{fig:SM_Transparency_effect}c (solid lines, left side) with no adjustable parameters. We find a reasonable agreement with the observed shifts and reproduce the expected saturation of the light shift due to the presence of a dark state. An optimized fit can be obtained with a slightly higher intensity corresponding to a waist of $\SI{21}{\micro \meter}$ instead of $\SI{23}{\micro \meter}$, and a slightly modified polarization distribution. In this fitted polarization distribution, the contribution of the weakest component, $\sigma^-$, is enhanced by a factor of roughly $2.5$. Both differences can be explained by effects from the vacuum chamber viewports and dielectric mirrors. 

When the left-hand circular polarization component of the transparency beam is added, we observe in Fig.~\ref{fig:SM_Transparency_effect}c (right side) that the ``dark" state shifts linearly away. In this manner all sub-levels of $^3\mathrm{P}_1$ can be shifted by more than $\SI{4}{\mega \hertz}$, more than $500$ times the linewidth of the laser cooling transition. For comparison, the light shift on the $^1\mathrm{S}_0$ ground state from the transparency beam is $\SI{20}{\kilo \hertz}$, and at most $\SI{380}{\kilo \hertz}$ by all trapping beams, about one order of magnitude smaller than the shift on $^3\mathrm{P}_1$ states from the transparency beam.

We demonstrate the protection achieved by the transparency beam in two ways. Firstly, we measure the lifetime of a pure BEC inside the dimple in the presence of all light and magnetic fields used in the CW BEC experiments. This pure BEC is produced beforehand using time-sequential cooling stages. Once the pure BEC is produced, we apply the same conditions as used for the CW BEC, except that the light addressing the \BlueMotTransition~transition is off, to prevent new atoms from arriving. Without the transparency beam the $1/e$ lifetime of a pure BEC in the dimple can't even reach $\SI{40}{\milli \second}$ while with the transparency beam it exceeds $\SI{1.5}{\second}$. 

\begin{figure}[tb]
\centerline{\includegraphics[width=0.98\columnwidth]{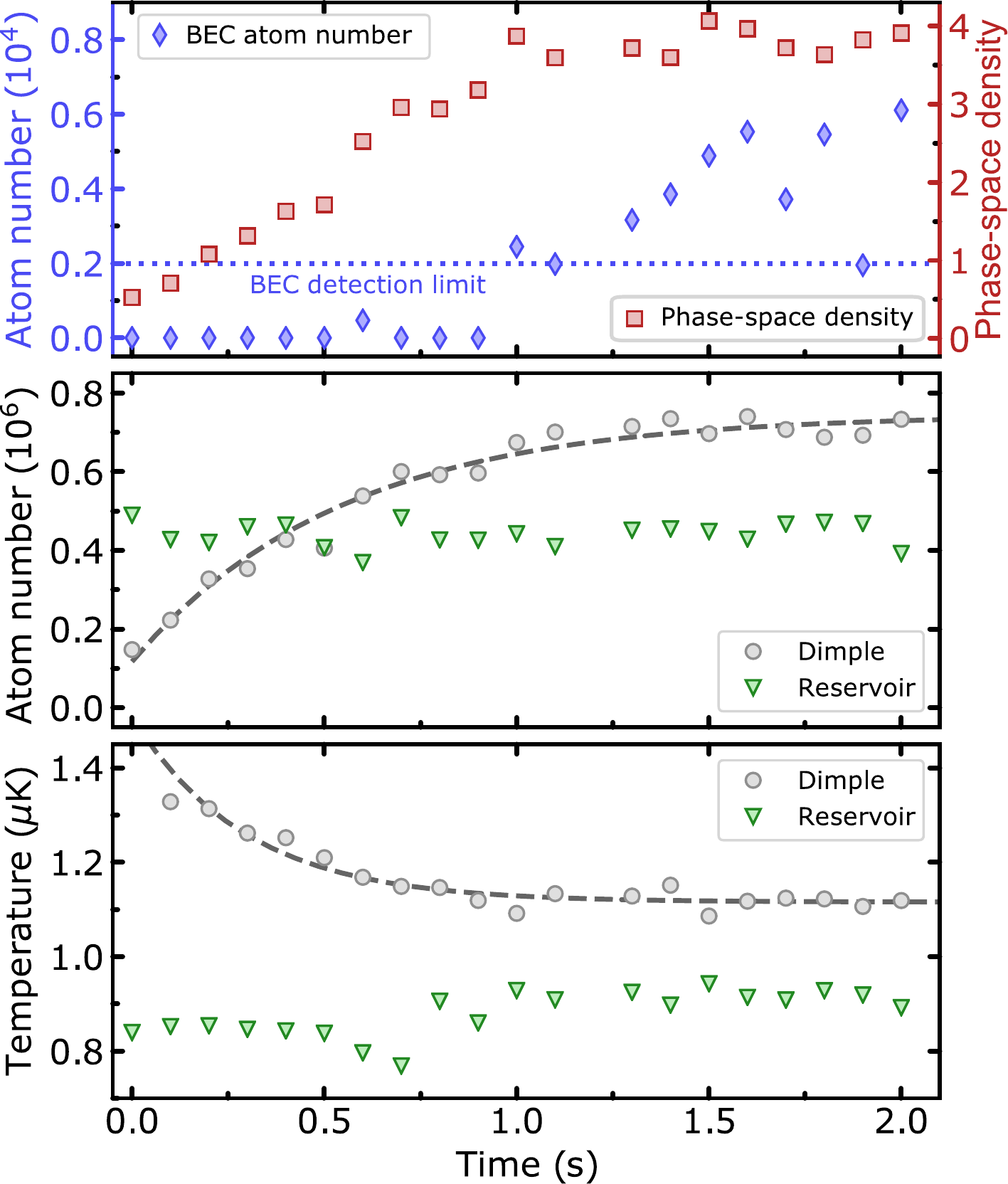}}
\caption{\label{fig:SM_Without_With_tranparency}\textbar \textbf{Influence of the transparency beam.} We let the experiment reach a steady state with the transparency beam off. At $t=0$ we switch the beam on and observe the system's evolution. We show (top) the BEC atom number and the phase-space density $\rho_D$ in the dimple. The blue dotted line indicates our BEC detection limit in terms of condensed atom number. We show (middle) the dimple and reservoir atom number, and (bottom) the temperature $T_D$ in the dimple and the temperature $T_{Ry}$ in the reservoir along the vertical axis. Both atom number and temperature in the reservoir remain constant, while the dimple loads additional atoms, indicating lower losses thanks to the protecting effect of the transparency beam.}
\end{figure}

Secondly, we show the influence of the transparency beam on the existence of a CW BEC. Beginning with the same configuration as the CW BEC but without the transparency beam, steady state is established after a few seconds, with no BEC formed. We then suddenly switch the transparency beam on and observe the sample's evolution as shown in Fig.~\ref{fig:SM_Without_With_tranparency}. While the reservoir sample seems unaffected, the dimple atom number increases by a factor of $6.4(1.8)$, indicating fewer losses. At the same time the sample (partially) thermalizes and a BEC appears after about $\SI{1}{\second}$. This demonstrates the critical importance of the transparency beam.

\medskip
\noindent
\textbf{Characterizing the BEC and thermal cloud} 

\noindent
To characterize the CW BEC and surrounding thermal cloud, we switch all traps and beams off, and perform absorption imaging. Fitting the expanding clouds' distributions allows us to estimate atom numbers and temperatures throughout the system, as well as the number of condensed atoms, all from a single image.

\begin{figure}[tb]
\centerline{\includegraphics[width=0.98\columnwidth]{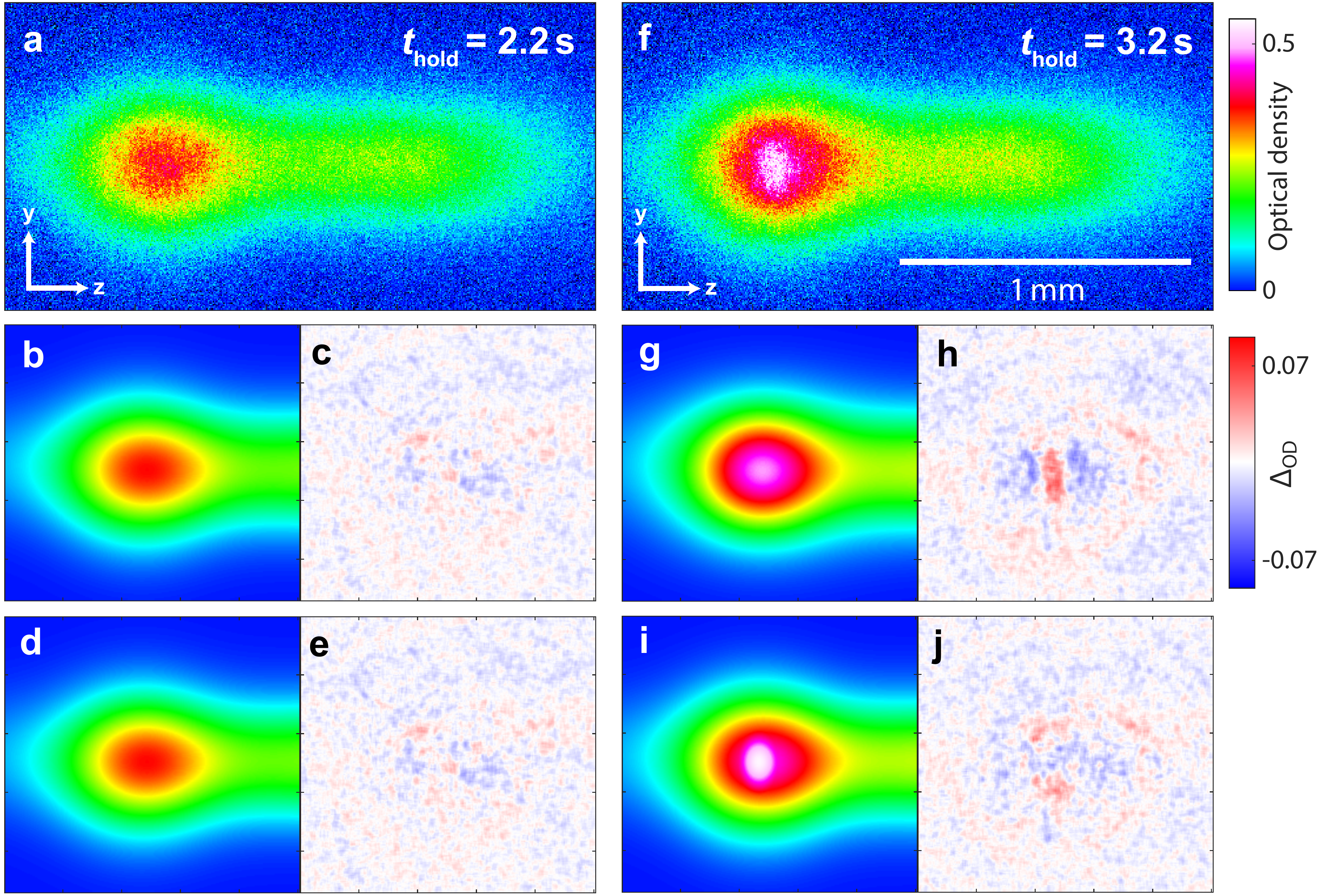}}
\caption{\label{fig:SM_Fitting}\textbar \textbf{Fitting a CW BEC.} We show absorption pictures and their respective fits for two hold times, before (left, $t_{\mathrm{hold}}= \SI{2.2}{\second}$) and after (right, $t_{\mathrm{hold}}= \SI{3.2}{\second}$) the formation of the BEC. In the top row, we show absorption pictures \textbf{a, f} taken after $\SI{18}{\milli \second}$ time-of-flight expansion. In the middle row, we show results of fits \textbf{b, g} to these pictures, and the fit residuals \textbf{c, h}. The fits use a 2D density distribution fit function accounting only for a thermal cloud. By contrast, the bottom row shows both fits \textbf{d, i} and residuals \textbf{e, j} with a 2D density distribution fit function including a Thomas-Fermi distribution describing a BEC, in addition to the thermal distribution. In presence of a BEC, the residual of the thermal-only fit \textbf{h} clearly shows a discrepancy at the BEC location, while the residual \textbf{j} demonstrates that the fit accounts for the BEC.}
\end{figure}

We begin with absorption images typically recorded after an $\SI{18}{\milli \second}$ time-of-flight expansion. The observed 2D density distribution can be fitted by an ensemble of four thermal components plus an additional Thomas-Fermi distribution when a BEC is present. Three independent 2D Gaussian functions represent atoms originating from the dimple, the reservoir, and the crossing between the guide and reservoir. Atoms originating from the guide are represented along the guide's axis by a sigmoid that tapers off due to the effect of the Zeeman slower, and in the radial direction by a Gaussian profile. Examples are shown in Fig.~\ref{fig:SM_Fitting}.

We found this fit function with $18$ free parameters to be the simplest and most meaningful one capable of representing the data. By combining knowledge of their distinct locations and/or momentum spreads, we can determine individually the populations and their characteristics. We find that the uncertainty in the fitted parameters is mostly insignificant compared to shot-to-shot variations in the data. An exception is distinguishing the population in the reservoir from the one in the guide-reservoir crossing region, where there is some ambiguity resulting in higher uncertainties. Both in the main text and these Materials and methods, the error bars indicate the standard deviation $\sigma$ calculated from multiple images. While it is possible to estimate the temperatures in the $y$ axis from a single fitted image, the initial cloud sizes in the $z$ direction are large compared to the ballistic expansion. Thus we use a set of measurements with varying times of flight to estimate $z$ axis temperatures.

When a BEC is present, it is necessary to add a Thomas-Fermi profile to the previously discussed fit function. The only additional free parameter used in the fit is the number of atoms in the BEC. We assume that the BEC position is the same as the one of the non-condensed atoms in the dimple and we calculate the BEC's radii from the BEC atom number, the \textit{s}-wave scattering length, the trap frequencies in the dimple, and the expansion time \cite{Castin1996BECTimeTrap}. These frequencies are calculated from the knowledge of the waists of each relevant beam and of the powers used. The waists are either directly measured or extracted from observations of dipole oscillation frequencies of a pure BEC in the trap for several beam powers. 

Adding an additional fitting parameter can lead to overfitting. To rigorously determine whether including this Thomas-Fermi distribution provides a significantly better fit of the data, we employ a statistical \textit{F}-test. This allows us to determine a BEC atom number threshold above which the fit is statistically better than the one without the Thomas-Fermi distribution. For this \textit{F}-test we isolate a region of interest (ROI) in the image containing both thermal and BEC atoms. We then calculate the value $F = \frac{(RRS_1 - RSS_2)}{p_2-p_1} / \frac{RSS_2}{n-p_2}$, where $RRS_i$ is the residual sum of squares over the ROI for model $i$ with $p_i$ parameters, and $n$ is the number of pixels of the ROI. The fit including the Thomas-Fermi distribution is significantly better than the one without, only if $F$ is higher than the critical value of an \textit{F}-distribution with $(p_2-p_1,n-p_2)$ degrees of freedom, with a desired confidence probability. By applying this test to the data of Fig.~\ref{fig:Steady_state_proof}, we find that the BEC model fits better, with a confidence greater than $\SI{99.5}{\percent}$, when the BEC atom number exceeds $2000$. This sets our detection limit, above which we are confident a BEC exists. Importantly, this limit is lower than the BEC atom number corresponding to a $- 2 \sigma_{\mathrm{N}}$ shot-to-shot fluctuation. This indicates that, at all times after steady state is reached, a BEC exists.

\medskip
\noindent
\textbf{Characterizing the reservoir loading}

\noindent
To estimate the flux $\Phi_{\mathrm{R}}$ of atoms loaded into the reservoir we begin with all laser cooling and trapping beams on, except for the Zeeman slower beam. After some time, a steady state is reached, in which the guide is filled but not the reservoir. We then switch on the Zeeman slower beam, and observe the loading of the reservoir, see Fig.~\ref{fig:SM_Loading_at_constant_flux}. Using our fitting procedure for the absorption images (see section ``Characterizing the BEC and thermal cloud"), we estimate the growth in the number of atoms in various regions of the cloud. With the rough assumptions of a constant flux $\Phi_{\mathrm{R}}$ and a one-body loss rate parameter $\Gamma_{\mathrm{loss}}$ originating for example from transfer of atoms to the dimple, the loading of the reservoir can be fitted by the exponential growth function $N_{\mathrm{R}}(t) = (1- e^{- \Gamma_{\mathrm{loss}} t}) \, \Phi_{\mathrm{R}} / \Gamma_{\mathrm{loss}}$. This function fits best for a flux $\Phi_{\mathrm{R}} = 1.4(2) \times 10^6 \, \mathrm{atoms}\, \si{\per \second}$, see Fig.~\ref{fig:SM_Loading_at_constant_flux}. We also show the combined number of atoms loaded in the reservoir and ``crossing" (guide-reservoir intersection) regions. We fit this data with a similar exponential growth function, and obtain a flux of $2.9(3) \times 10^6 \, \mathrm{atoms}\, \si{\per \second}$. The relative uncertainty of the combined atom numbers is smaller than for the data set describing only the reservoir atom number. This is due to ambiguity between the ``reservoir" and ``crossing" regions in our fitting procedure.

\medskip
\noindent
\textbf{Loading dynamics}

\noindent
The data of Fig.~\ref{fig:SM_Loading_at_constant_flux} show several timescales at play in the system's evolution. First we see atoms populating the reservoir. About $\SI{500}{\milli \second}$ later the dimple region begins to fill. Finally about $\SI{1}{\second}$ after the start of the dimple loading we see a BEC begin to form. 

These dynamics can be understood from the thermalization timescale and from the need to exceed the critical phase-space density to begin forming a BEC. We estimate the peak phase-space density evolution by $\rho_D = N_D \left(\frac{\hbar^3 \omega_{Dx} \omega_{Dy} \omega_{Dz}}{k_B^3 T_{D}^3}\right)$, where $N_D$ is the thermal atom number in the dimple. However, this estimation is inaccurate because of the non-thermalized distribution function describing the atoms in the dimple (see section ``Modeling of the BEC's open dynamics" below). We give it here as an indication of the phase-space density, which with this definition and for a thermalized sample should be greater than $1.2$ in order to produce a BEC.

Once the critical phase-space density is exceeded BEC formation begins at a slow rate and then accelerates as more atoms condense. Indeed, the growth of a BEC or matter-wave is governed by Bose-stimulated scattering, which scales with the number of atoms occupying the ground state of the trap \cite{Borde1995AtomStiEmiss, Miesner1998BoseEnhancedBEC, Kozuma1999AmplMatterWave, Inouye1999AmplMatterWave}.

After about $\SI{3}{\second}$ steady state is established. We probe the system for various hold times up to $\SI{5}{\minute}$ and find no indication that the BEC departs from steady state. This is exemplified in Fig.~\ref{fig:SM_Loading_at_constant_flux} and Fig.~\ref{fig:Steady_state_proof}b.

\medskip
\noindent
\textbf{BEC anisotropy after time-of-flight} 

\noindent
A Bose-Einstein condensate and a thermal gas expand very differently when released from a trap. A thermal gas expands isotropically, beginning with a shape reminiscent of the initial trap geometry, and asymptotically approaching an isotropic shape for long times of flight (TOF). By contrast, the expansion of a BEC released from an anisotropic trap remains anisotropic for long TOF, inverting its aspect ratio mid flight. This originates from the anisotropic release of mean-field energy, which reflects the trap anisotropy \cite{Castin1996BECTimeTrap}.

\begin{figure*}[tb]
\centerline{\includegraphics[width=0.65\textwidth]{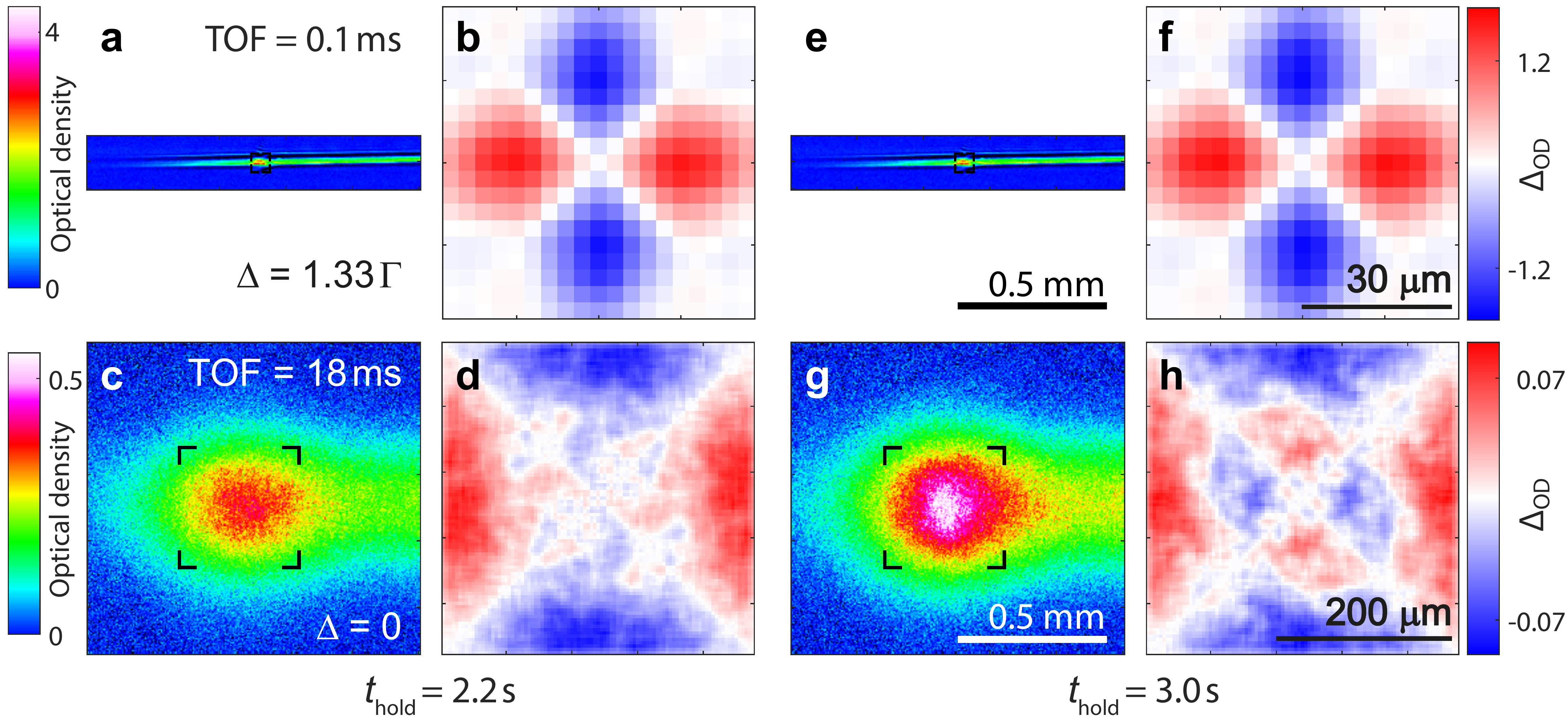}}
\caption{\label{fig:SM_Anisotropy}\textbar \textbf{BEC anisotropy after time-of-flight.} \textbf{a, e,} Absorption images after short ($\SI{0.1}{\milli \second}$) and \textbf{c, g,} after long ($\SI{18}{\milli\second}$) free-expansion time-of-flight (TOF), for short ($t_{\mathrm{hold}} = \SI{2.2}{\second}$, \textbf{a, c}) and long ($t_{\mathrm{hold}} = \SI{3.0}{\second}$, \textbf{e, g}) trap loading times. Pictures \textbf{a} and \textbf{e} were imaged with a detuning of $1.33 \, \Gamma$ to avoid saturation. The regions of interest (corner-marked squares) centered around the density maximums are analyzed in panels \textbf{b, d, f, h}, which show the transpose-anisotropy of the density distribution. This representation produces a cloverleaf pattern when the atomic cloud is anisotropic (see text). For short $t_{\mathrm{hold}}$ (left), the cloverleaf pattern, which appears because of how the trap geometry initially shapes the thermal cloud, keeps a constant sign and diminishes during the expansion of a thermalized gas sample. For long $t_{\mathrm{hold}}$ (right), we observe at long expansion time \textbf{h} an additional smaller cloverleaf pattern of opposite sign, which is indicative of the presence of a BEC.}
\end{figure*}

The transpose-anisotropy provides an elegant method by which to efficiently detect the presence of a CW BEC. The transpose-anisotropy of the density distribution $n_{\mathrm{OD}}$ is $n^s_{\mathrm{OD}}\left(y, z\right)- n^s_{\mathrm{OD}}\left(z, y\right)$, where the origin of the coordinate system is at the density maximum. $n^s_{\mathrm{OD}}$ is obtained from $n_{\mathrm{OD}}$ by adding the same $n_{\mathrm{OD}}$ distribution rotated by $\SI{180}{\degree}$. Transpose-anisotropies for short ($\SI{0.1}{\milli \second}$) and long ($\SI{18}{\milli\second}$) TOFs are shown in Fig.~\ref{fig:SM_Anisotropy}. For $\SI{0.1}{\milli \second}$ TOF, the density distribution shows a marked anisotropy as indicated by a strong cloverleaf pattern. This initial anisotropy is solely due to the action of the trap geometry on the density distribution of the thermal gas, as the size of a potential BEC is below our imaging resolution. However, for $\SI{18}{\milli\second}$ TOF we see a difference between pictures for short ($\SI{2.2}{\second}$) and long ($\SI{3.0}{\second}$) hold time $t_{\mathrm{hold}}$. For short $t_{\mathrm{hold}}$ the anisotropy is broad and simply a remnant of the initial cloud anisotropy. However for long $t_{\mathrm{hold}}$, at which steady state is established, we see an additional, smaller cloverleaf pattern with opposite anisotropy around the center of the picture. Both the existence and the sign of this pattern are consistent with the expansion of a BEC from a dimple with our trap frequencies.

\medskip
\noindent
\textbf{Modeling of the BEC's open dynamics}

\noindent
In order to describe the formation and properties of the continuous-wave Bose-Einstein condensate, we develop a model capable of fitting our experimental data, as shown for example by the dashed blue line in Fig.~\ref{fig:Steady_state_proof}b. In the following we describe the model and its results.

Firstly, we lay out the model, which builds on the standard Boltzmann kinetic theory of Bose gases with additional dissipation to account for the open dynamics. As a main result of our derivation we arrive at the rate equation Eq.~(\ref{eq:Nrate}). We provide details of the derivation of each gain and loss term in the equation and the assumptions that led to them.

Secondly, we discuss the fit of this model to our data, from which we extract useful physical quantities, like the 1- and 3-body loss rate parameters, and the flux into the CW BEC after reaching steady state.  

In particular, we find that we are not able to describe our data if we assume that our system is in thermal equilibrium. This reflects the intrinsic driven-dissipative nature of the system.

~\\
\textit{The model} --- We model the dynamics of the BEC using the rate equation

\beqa
	\dot{n}_{\rm BEC}({\bf r}) &=& s_{\rm in}({\bf r})-s_{\rm out}({\bf r}) - \gamma_{\rm 1b} n_{\rm BEC}({\bf r}) \nonumber \\
	&& - \gamma_{\rm 3b} \left[ n_{\rm BEC}({\bf r})^3 + 6 n_{\rm BEC}({\bf r})^2 n_{\rm th}({\bf r}) \right. \nonumber \\
	&& \phantom{ahoj} \left. + 6 n_{\rm BEC}({\bf r}) n_{\rm th}({\bf r})^2 \right], \nonumber \\
	\label{eq:rate_dens_2}
\eeqa
where $\gamma_{\rm 1b},\gamma_{\rm 3b}$ are the phenomenological one- and three-body loss rate parameters, see \cite{Dutta2015BECgrowthDimple}, $n_{\rm BEC}, n_{\rm th}$ are the local densities of the BEC and thermal atoms in the dimple, and we do not write explicitly the time dependence of the variables. In the framework of kinetic Boltzmann equations, $s_{\rm in}, s_{\rm out}$ are the collisional integrals (source terms) describing the exchange between the thermal atoms and the BEC. To this end we follow closely the treatment in \cite{Jackson2002TheoBECfiniteT} and write (setting ${\bf v}_{\rm BEC}=0$ in \cite{Jackson2002TheoBECfiniteT})
\begin{subequations}
	\label{eq:s_in_out}
	\begin{align}		
		s_{\rm in} &= \frac{n_{\rm BEC} \sigma}{\pi h^3} \int {\rm d} {\bf p}_3 f_3 \frac{1}{v_3} \int {\rm d} {\bf \tilde{v}} f_4 \label{eq:s_in} \\
		s_{\rm out}	&= \frac{n_{\rm BEC} \sigma}{h^3} \int {\rm d} {\bf p}_2 f_2 \Delta v \int \frac{{\rm d}\Omega}{4\pi} (1+f_3+f_4), \label{eq:s_out}
	\end{align}
\end{subequations}
where $\sigma=8\pi a_{\rm sc}^2$ is the \textit{s}-wave scattering cross-section, $\Delta v = \sqrt{v_2^2 - 4 v_0^2}$, $v_0=\sqrt{g n_{\rm BEC}/m}$, $g=4\pi \hbar^2 a_{\rm sc}/m$ and ${\bf p}_j = m {\bf v}_j$. Eq.~(\ref{eq:s_in}) describes the scattering of two thermal atoms with velocities ${\bf v}_3$ and ${\bf v}_4$ resulting in a BEC atom and a thermal atom with velocity ${\bf v}_2$; Eq.~(\ref{eq:s_out}) corresponds to the opposite process. In Eq.~(\ref{eq:s_in}) ${\bf v}_4 = {\tilde {\bf v}} + \frac{g n_{\rm BEC}}{m v_3^2}{\bf v}_3$ with $\tilde{\bf v} \perp {\bf v}_3$ such that the second integration is performed over the plane perpendicular to ${\bf v}_3$. In Eq.~(\ref{eq:s_out}) ${\rm d}\Omega$ denotes the solid angle subtended by ${\bf v}_3$ and ${\bf v}_4$. We use \Sr{84} atomic mass $m$ and its \textit{s}-wave scattering length $a_{\rm sc} = 122.8 \, a_0$, with $a_0$ the Bohr radius. $f_j = f({\bf r},{\bf p}_j)$ are the \textit{unknown} distribution functions of the thermal atoms with the property $N_{\rm th} = \int{\rm d}{\bf r} \, {\rm d}{\bf p}/h^3 f({\bf r},{\bf p})$. 

In principle, one could obtain $f_j$ from complete N-body simulations \cite{Jackson2002TheoBECfiniteT} of the open system, accounting for the coupled dynamics between the reservoir, the dimple and the BEC. While strongly desirable, such a study goes beyond the scope of the present analysis\footnote{In this context, a possibility is to consider instead a simpler ergodic description of the Bose-Einstein condensation \cite{Luiten1996KineticEvapCool, Gardiner1998BECgrowth, Holland1997BECkinetic}. However, our preliminary analysis indicates breaking of the ergodicity assumptions, thus requiring the complete N-body simulations.}. Instead, let us first assume a thermal equilibrium such that $f$ is given by the Bose-Einstein distribution. In this case the total number of non-interacting atoms in a harmonic trap is given by $N_{\rm th} = (\beta_D \hbar \omega_D)^{-3} {\rm Li}_3\left({\rm e}^{-\beta_D \mu} \right)$. Here $\mu$ is the chemical potential of the distribution, $\omega_D = (\omega_{Dx}\omega_{Dy}\omega_{Dz})^{1/3}$, $\beta_D = 1/k_{\rm B}T_D$, and ${\rm Li}_s(z)$ is a polylogarithm of order $s$. With the experimentally measured dimple trap frequencies and gas temperature, we find that even for the maximum allowed $\mu = 0$, the atom number only reaches $N_{\rm th} \approx 1.8 \times 10^5$. This estimation is well below the measured value of $6.9(4) \times 10^5$ atoms in the dimple. To try to resolve this discrepancy, we perform a more refined calculation following \cite{Stellmer2013LaserCoolingToBEC}, where the harmonic trapping potential is replaced by the combined potential of the reservoir and dimple traps, and where we account for the interactions between atoms. This calculation gives $N_{\rm th} \approx 3.6 \times 10^5$ atoms in reservoir and dimple combined, which is less than those measured in the dimple, let alone in the reservoir and dimple combined ($1.4(2) \times 10^6$). Unsurprisingly, in this calculation most atoms are contained well within the dimple, as $k_{\rm B} T_D$ is much smaller than the dimple depth. The refined approach is thus also incapable of reproducing the measured atom number. This is in direct contradiction with our previous analysis in \cite{Stellmer2013LaserCoolingToBEC}. There, the refined calculation was capable of accounting for all atoms in a trap with a similar geometry, but where the continuous refilling of the reservoir was absent. This difference leads us to conclude that the current system does not fulfill the thermal equilibrium assumption due to its open character, and that the population dynamics are governed by out-of-equilibrium distribution functions $f$.

To proceed, we assume that the distribution function can be written as a product 
\beq
    f({\bf r},{\bf p}) = n_{\rm th}({\bf r}) f'({\bf p}). \label{eq:f_assumption}
\eeq
We take $f'({\bf p}) = \frac{1}{\cal N}{\rm e}^{-\beta_D \frac{p^2}{2m}}$, ${\cal N} = \frac{4\pi}{(2\pi\hbar)^3}\sqrt{\frac{\pi}{2}}\left( \frac{m}{\beta_D} \right)^\frac{3}{2}$ to be the Boltzmann distribution, which fulfills the normalization property $\int \frac{{\rm d}^3 p}{(2\pi\hbar)^3} f'({\bf p}) = 1$. We next specify the densities $n_{\rm BEC}({\bf r})$ of the BEC and $n_{\rm th}(\bf r)$ of the thermal atoms appearing in Eq.~(\ref{eq:rate_dens_2}).
For the BEC we consider the Thomas-Fermi profile given by
\beqa
	n_{\rm BEC}({\bf r})= n_0 \left(1-\left(\frac{x}{R_x}\right)^2 -\left(\frac{y}{R_y}\right)^2 -\left(\frac{z}{R_z}\right)^2\right), \nonumber \\
	\label{eq:nBEC_profile}
\eeqa
where $R_\alpha = \sqrt{2 \mu_D/m}/{\omega_{D \alpha}}$, $\mu_D = \frac{\hbar \omega_D}{2} \left( 15 N_{\rm BEC} \frac{a_{\rm sc}}{a_{\rm ho}} \right)^\frac{2}{5} $, $a_{\rm ho} = \sqrt{\frac{\hbar}{m \omega_D}}$, $N_{\rm BEC}$ is the total BEC atom number and $n_0 = \mu_D/g$. In the BEC region, the condensate atoms will repel the thermal ones resulting, to a good accuracy, in the characteristic parabolic profile
\beqa
	n_{\rm th}({\bf r}) \approx \nonumber \\ 
	& n_{{\rm th},0} \left[ 1 + \gamma_x \left(\frac{x}{R_x}\right)^2 + \gamma_y \left(\frac{y}{R_y}\right)^2 + \gamma_z \left(\frac{z}{R_z}\right)^2 \right], \nonumber \\
	\label{eq:nth_profile}
\eeqa
where $\gamma_\alpha = (n_{c,\alpha}-n_{{\rm th},0})/n_{{\rm th},0}$.
Next, motivated by the approach in \cite{Stellmer2013LaserCoolingToBEC, Jackson2002TheoBECfiniteT}, we determine the thermal densities in the center, $n_{{\rm th},0}=n_{{\rm th}}(0)$, and at the edge of the BEC cloud, $n_{c,\alpha}=n_{{\rm th}}(R_\alpha)$ in the direction $\alpha=x,y,z$, self-consistently as
\beqa
	n_{\rm th}({\bf r}) = -\frac{1}{\lambda_{\rm dB}^3} {\rm Li}_\frac{3}{2} \left[ -{\rm e}^{-\beta_D (V({\bf r}) + 2gn_{\rm BEC}({\bf r}) + 2g n_{\rm th}({\bf r})-\eta)} \right], \nonumber \\
	\label{eq:nth_self_FD}
\eeqa
with $\lambda_{\rm dB} =h/\sqrt{2\pi m k_B T_D}$ the thermal de-Broglie wavelength. Here, we need to stress that unlike in \cite{Stellmer2013LaserCoolingToBEC}, the function in Eq.~(\ref{eq:nth_self_FD}) is not the density corresponding to the thermal equilibrium of a Bose gas as that one fails to reproduce the observed number of thermal atoms $N_{\rm th}$, see the discussion above Eq.~(\ref{eq:f_assumption}). The function in Eq.~(\ref{eq:nth_self_FD}) in fact corresponds to a Fermi-Dirac distribution, which however here does not have a particular physical significance. It should be seen as a convenient ansatz from which we extract $n_{{\rm th},0}$ and $n_{c,\alpha}$, which has the advantage that it allows to match the observed number $N_{\rm th}$ for positive values of the parameter $\eta$ \cite{Dingle1957FermiIntegral}.

The assumed prescription for $n_{\rm th}({\bf r})$ Eq.~(\ref{eq:nth_self_FD}) is clearly an idealization. Despite this fact, the model yields a value for the three-body loss rate parameter $\gamma_{\rm 3b}$ that is compatible with the values reported in the literature (see section \textit{The results} below). We interpret this compatibility as indicating that the values of $n_{\rm th,0}$ and $n_{c,\alpha}$ extracted from Eq.~(\ref{eq:nth_self_FD}) lie within a factor $O(1)$ from the actual experimental values.

Next, we turn to the evaluation of the collisional integrals Eqs.~(\ref{eq:s_in_out}) using the distribution function Eq.~(\ref{eq:f_assumption}) with the result
\begin{subequations}
	\begin{align}
		\label{eq:s_in_out_2}
		s_{\rm in} &= n_{\rm BEC} n_{\rm th}^2 \sigma I_{\rm in} \\
		s_{\rm out} &= n_{\rm BEC} n_{\rm th} \sigma I_{\rm out}^{(1)} + n_{\rm BEC} n_{\rm th}^2 \sigma I_{\rm out}^{(2)}.
	\end{align}
\end{subequations}
Here, we have introduced the functions
\begin{subequations}
	\label{eq:I_in_out}
	\begin{align}		
		I_{\rm in} &= \kappa v_0^2 K_1(\beta_D m v_0^2) \\
		I_{\rm out}^{(1)} &= \kappa {\cal N} {\rm e}^{-\beta_D m v_0^2} v_0^2 K_1(\beta_D m v_0^2) \\
		I_{\rm out}^{(2)} &= 2 \kappa j,
	\end{align}
\end{subequations}
where $\kappa=\frac{1}{h^3} \frac{1}{{\cal N}^2} \frac{8 \pi m^3}{m \beta_D}$, $K_1$ is the modified Bessel function of the second kind, and 
\begin{widetext}
\begin{equation}
	j({\bf r}) = \int_{2 v_0}^\infty {\rm d}v_2 \, v_2 {\rm e}^{-\frac{\beta_D m}{2}v_2^2} \left[ {\rm e}^{-\frac{\beta_D m}{8}(v_2 - \Delta v(v_2))^2} - {\rm e}^{-\frac{\beta_D m}{8}(v_2 + \Delta v(v_2))^2} \right] ,
\end{equation}
\end{widetext}
which has to be evaluated numerically.

Ultimately, we are interested in the dynamics of the total number of the condensate atoms, which can be obtained by integrating Eq.~(\ref{eq:rate_dens_2}) over the volume. To be able to carry out the final integration, we take the functions $I_{\rm in}, I_{\rm out}^{(1,2)}$ to be independent of the position\footnote{In principle, they still depend on spatial coordinates through $v_0(n_{\rm BEC}({\bf r}))$. For the span of experimental values of $n_{\rm BEC}({\bf r})$, this leads to a maximum variation by a factor $0.7$.}. Using the profiles (\ref{eq:nBEC_profile}),(\ref{eq:nth_profile}) we arrive at the final rate equation
\beq
	\dot{N}_{\rm BEC} = S_{\rm in} - S_{\rm out} -L_{\rm 1b}- L_{\rm 3b}^{(3)} - L_{\rm 3b}^{(2)} - L_{\rm 3b}^{(1)},
	\label{eq:Nrate}
\eeq
where
\begin{subequations}
	\label{eq:Nrate_terms}
	\begin{align}
		S_{\rm in} &= n_{0} n_{\rm th,0}^2 \sigma V^{(1,2)} I_{\rm in} \label{eq:Sin} \\
		S_{\rm out} &= n_{0} n_{\rm th,0} \sigma V^{(1,1)} I_{\rm out}^{(1)} + n_{0} n_{\rm th,0}^2 \sigma V^{(1,2)} I_{\rm out}^{(2)} \label{eq:Sout}\\
		L_{\rm 1b} &= \gamma_{\rm 1b} N_{\rm BEC} \label{eq:L1b}\\
		L_{\rm 3b}^{(3)} &= \gamma_{\rm 3b} n_0^3 V^{(3,0)} \label{eq:L3b_3} \\				
		L_{\rm 3b}^{(2)} &= 6 \gamma_{\rm 3b} n_0^2 n_{\rm th,0} V^{(2,1)} \label{eq:L3b_2} \\
		L_{\rm 3b}^{(1)} &=6 \gamma_{\rm 3b} n_0 n_{\rm th,0}^2 V^{(1,2)}	\label{eq:L3b_1}	
   \end{align}
\end{subequations}
In the above equations, we have introduced the volume integrals 
\begin{equation}
V^{(p,q)}=1/ (n_0^p n_{\rm th,0}^q) \int {\rm d}{\bf r} n_{\rm BEC}({\bf r})^p n_{\rm th}({\bf r})^q,
\end{equation}
which evaluate to
\begin{subequations}
	\begin{align}
		V^{(1,1)} &= 4\pi R_x R_y R_z \frac{2}{105}(7+3\bar{\gamma}) \\
		V^{(1,2)} &= 4\pi R_x R_y R_z \frac{2}{315}(21+5G+18\bar{\gamma}) \\
		V^{(2,1)} &= 4\pi R_x R_y R_z \frac{8}{315}(3+\bar{\gamma}) \\
		V^{(3,0)} &= 4\pi R_x R_y R_z \frac{16}{315},
   \end{align}
   \label{eq:Vpq}
\end{subequations}
where
\begin{subequations}
	\begin{align}
		\bar{\gamma} &= \frac{1}{3}(\gamma_x + \gamma_y + \gamma_z) \\
		G &= \frac{1}{5} (\gamma_x^2 + \gamma_y^2 + \gamma_z^2) + \frac{2}{15}(\gamma_x \gamma_y + \gamma_x \gamma_z + \gamma_y \gamma_z).
	\end{align}
\end{subequations}
To fit the data, we take $\gamma_{\rm 1b}, \gamma_{\rm 3b}$ in Eqs.~(\ref{eq:L1b})-(\ref{eq:L3b_1}) as free parameters and further parametrize Eq.~(\ref{eq:Sin}) and Eq.~(\ref{eq:Sout}) as
\begin{subequations}
    \label{eq:S_in_out_fit}
	\begin{align}
    S_{\rm in} &= \alpha_{\rm in} n_{0} n_{\rm th,0}^2 \sigma V^{(1,2)} I_{\rm in} \label{eq:Sin_fit} \\
	S_{\rm out} &= \alpha_{\rm out} \alpha_{\rm in} (n_{0} n_{\rm th,0} \sigma V^{(1,1)} I_{\rm out}^{(1)} + n_{0} n_{\rm th,0}^2 \sigma V^{(1,2)} I_{\rm out}^{(2)}) \label{eq:Sout_fit}
	\end{align}
\end{subequations}
with $\alpha_{\rm in}, \alpha_{\rm out}$ the fit parameters.

\begin{figure}[H]
\centerline{\includegraphics[width=0.98\columnwidth]{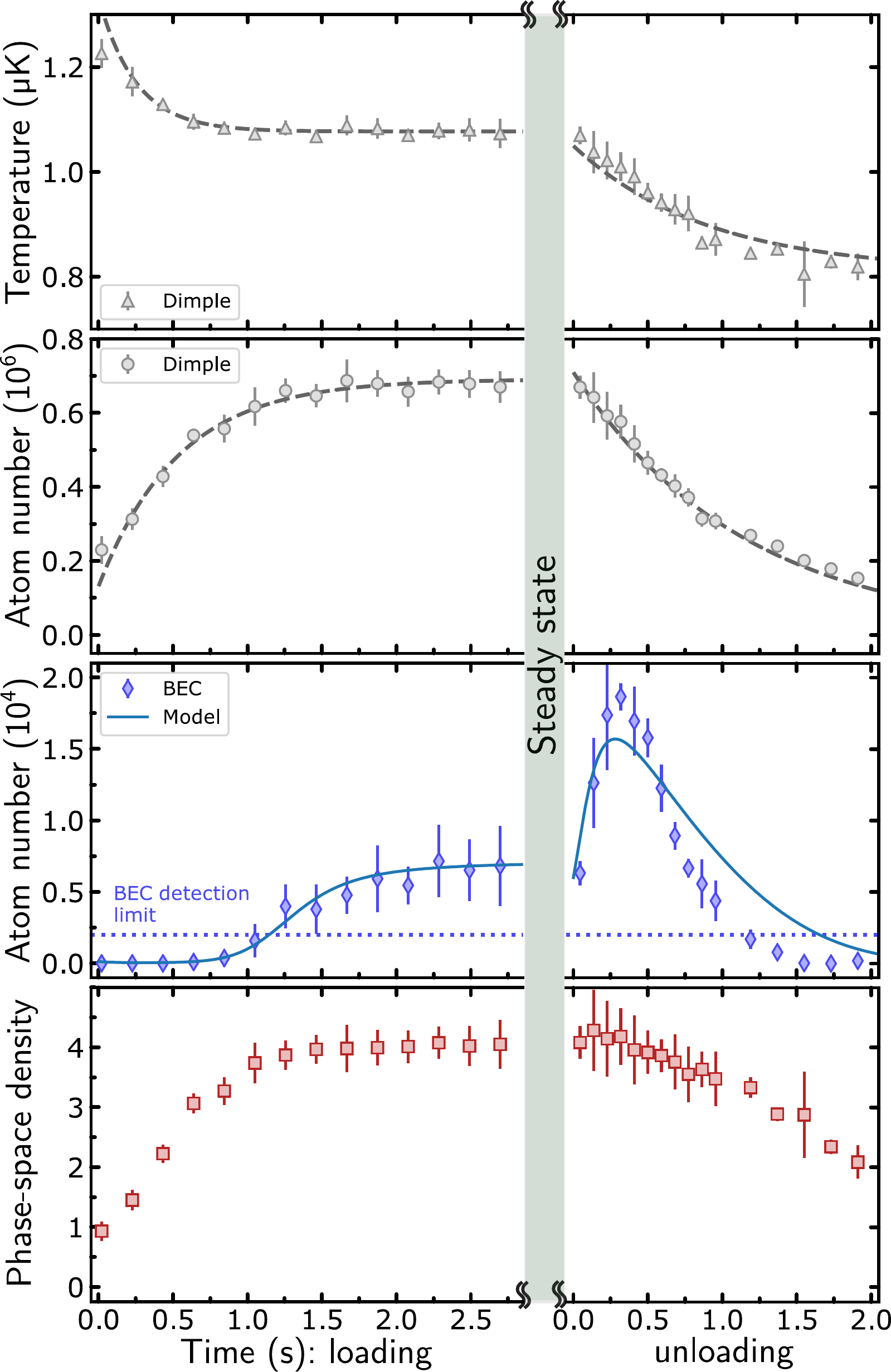}}
\caption{\label{fig:SM_Loading_unloading_plus_model}\textbar \textbf{Modeling the onset and disappearance of the BEC.} Evolution of the gas in the dimple. (left) BEC onset when loading reservoir and dimple at constant flux $\Phi_{\mathrm{G}}$ in the guide, started by Zeeman slower switch-on at $t=\SI{-0.7}{\second}$ (same data as in Fig.~\ref{fig:SM_Loading_at_constant_flux}, but choosing a different origin for the time axis). (right) BEC disappearance after setting $\Phi_{\mathrm{G}}$ to zero by switching the Zeeman slower off. The first two rows show the temperature and atom number of the thermal atoms in the dimple. The last two rows show the BEC atom number and the estimated phase-space density $\rho_D$ in the dimple. The solid blue line is the result of the fit of the BEC atom number with the model of Eq.~(\ref{eq:Nrate}). The dashed grey lines are results of fits with exponential growth/decay functions, which are used as input for Eq.~(\ref{eq:Nrate}). Error bars represent one standard deviation $\sigma$ from binning $\sim 4$ data points. For the BEC atom number, due to the small number of data points, the error bars can be underestimated compared to the, more reliable, characterization at steady state of $\sigma_{\mathrm N} = 2300$ provided in the main text. The phase-space density $\rho_D$ is estimated from measurements of the atom number and temperature in the dimple (see text).}
\end{figure}

~\\
\textit{The results} --- We have simultaneously fitted $6$ data sets of the onset of the BEC analogous to Fig.~\ref{fig:Steady_state_proof}b, with slight variations in the experimental starting conditions. For each of these sets, we use as a known time-varying input our measurements of the time evolution of dimple atom number and temperature (see for example the exponential fits to the dimple data of Fig.~\ref{fig:SM_Loading_unloading_plus_model}). We use the rate Eq.~(\ref{eq:Nrate}) and require the fit parameters $\gamma_{\rm 1b}$, $\gamma_{\rm 3b}$, and $\alpha_{\rm out}$ to be the same for all the data sets. Conversely, we allow independent variation for each data set of  $\alpha_{\rm in}$ and the initial BEC atom number $N_{\rm BEC}(0)$, which provides the seed to Eq.~(\ref{eq:Nrate}). 

From this single fit, we obtain the loss rate parameters $\gamma_{\rm 1b} = \SI{6(2)}{\per \second}$ and $\gamma_{\rm 3b} = 1.9(2) \times 10^{-29} \, \si{\centi \meter^6 \per \second}$. The value obtained for $\gamma_{\rm 3b}$ is compatible with the rough experimental values reported in the literature, namely ${\approx}0.7 \times 10^{-29} \, \si{\centi \meter^6 \per \second}$ in \cite{MartinezdeEscobar2012PhDThesis} and $1.4(3) \times 10^{-29} \, \si{\centi \meter^6 \per \second}$. The latter value can be extracted from the data used in \cite{Stellmer2013ProdQuGasSr}, where the standard deviation is derived from statistical uncertainty, without accounting for systematic effects. We also obtain the average flux into the CW BEC at steady state $\bar{S}_{\rm in} - \bar{S}_{\rm out} = 2.4(5) \times 10^5 \, \mathrm{atoms}\, \si{\per \second}$. In steady state, most of this incoming flux is compensated by the three-body losses involving thermal atoms described by Eqs.~(\ref{eq:L3b_2}) and (\ref{eq:L3b_1}), while the loss mechanisms Eqs.~(\ref{eq:L1b}) and (\ref{eq:L3b_3}) are only reaching a few $10^4 \, \mathrm{atoms}\, \si{\per \second}$. Finally, we extract from the fit additional information about the CW BEC, namely the Thomas-Fermi radii $\{ R_{\mathrm{TFx}}, R_{\mathrm{TFy}}, R_{\mathrm{TFz}} \} = \{ 2.8(1), 1.2(1), 2.9(1) \} \, \si{\micro \meter}$, the peak BEC density $n_0 = 3.6(2) \times 10^{20} \,\mathrm{atoms}\,\si{\per \cubic \meter}$ and the density of the thermal gas at the BEC center $n_{\mathrm{th,0}} = 3.6(1) \times 10^{20} \,\mathrm{atoms}\,\si{\per \cubic \meter}$.   

Fig.~\ref{fig:SM_Loading_unloading_plus_model} shows example fits together with their corresponding experimental data sets. The left (``loading" stage) of Fig.~\ref{fig:SM_Loading_unloading_plus_model} shows the onset and stabilization of the BEC following the switch on of the Zeeman slower beam. The right (``unloading" stage) of Fig.~\ref{fig:SM_Loading_unloading_plus_model} shows the subsequent data after suddenly switching off the same beam. This corresponds to stopping the atomic flux into the reservoir and leads to the disappearance of the BEC. It is apparent that the rate equation Eq.~(\ref{eq:Nrate}) captures well the initial dynamics of the BEC atom number as well as the initial dynamics of the unloading stage. The discrepancy for longer times during the unloading stage can likely be attributed to the change of the momentum distribution functions $f({\bf r},{\bf p})$. This arises from the simultaneous thermalization and depletion of the dimple atoms, which is not captured by the parametrization of Eqs.~(\ref{eq:S_in_out_fit}).

In summary, the rate equation (\ref{eq:Nrate}) provides a satisfactory fit to the data, from which we obtain the loss rate parameters $\gamma_{\rm 1b}$, $\gamma_{\rm 3b}$ and the steady-state BEC flux rate $\bar{S}_{\rm in} - \bar{S}_{\rm out}$. A more comprehensive quantitative understanding may be provided through further analysis starting from first principles. Such an analysis requires more complex modeling of the coupled dynamics between the reservoir, the dimple, and the BEC, such as the use of the ZNG theory \cite{Zaremba1999DynamcsTrapBoseGas, Bijlsma2000TheoGrowthBEC} and the related $N$-body simulations \cite{Jackson2002TheoBECfiniteT}.

\end{document}